\documentclass[manuscript,screen]{acmart}
\AtBeginDocument{%
  \providecommand\BibTeX{{%
    \normalfont B\kern-0.5em{\scshape i\kern-0.25em b}\kern-0.8em\TeX}}}

\usepackage{balance}
\usepackage{marvosym}
\usepackage{tikz}
\usepackage{algorithm}
\usepackage{amsmath}
\usepackage{algpseudocode}
\usepackage{caption}
\usepackage{subcaption}

\usepackage{bbding}
\usepackage{bbding}
\algnewcommand\algorithmicforeach{\textbf{for each}}
\algdef{S}[FOR]{ForEach}[1]{\algorithmicforeach\ #1\ \algorithmicdo}
\algnewcommand\algorithmicto{\textbf{to}}
\algdef{SE}[FOR]{ForTo}{EndFor}[2]{%
  \algorithmicfor\ #1\ \algorithmicto\ #2\ \algorithmicdo
}{%
  \algorithmicend\ \algorithmicfor
}
\setcopyright{acmlicensed}

\begin{document}
\fancyhead{}
\title{A Meta-Evaluation of C/W/L/A Metrics}
\subtitle{System Ranking Similarity, System Ranking Consistency and Discriminative Power}
\author{Nuo Chen}
\email{pleviumtan@toki.waseda.jp}
\affiliation{%
  \institution{Waseda University}
  \city{Tokyo}
  \country{Japan}}

\author{Tetsuya Sakai}
\email{tetsuyasakai@acm.org}
\affiliation{%
  \institution{Waseda University}
  \city{Tokyo}
  \country{Japan}}
\begin{abstract}
Recently, Moffat \textit{et al.} proposed an analytic framework, namely C/W/L/A, for offline evaluation metrics. This framework allows information retrieval (IR) researchers to design evaluation metrics through the flexible combination of user browsing models and user gain aggregations. However, the statistical stability of C/W/L/A metrics with different aggregations is not yet investigated. In this study, we investigate the statistical stability of C/W/L/A metrics from the perspective of:~(1)~the system ranking similarity among aggregations,~(2)~the system ranking consistency of aggregations and~(3)~the \textit{discriminative power} of aggregations. More specifically, we combined various aggregation functions with the browsing model of Precision, Discounted Cumulative Gain~(DCG), Rank-Biased Precision~(RBP), INST, Average Precision (AP) and Expected Reciprocal Rank~(ERR), examing their performances in terms of system ranking similarity, system ranking consistency and discriminative power on two offline test collections. Our experimental result suggests that, in terms of system ranking consistency and discriminative power, the aggregation function of \textit{expected rate of gain} (ERG) has an outstanding performance while the aggregation function of \textit{maximum relevance} usually has an insufficient performance. The result also suggests that Precision, DCG, RBP, INST and AP with their canonical aggregation all have favourable performances in system ranking consistency and discriminative power; but for ERR, replacing its canonical aggregation with ERG can further strengthen the discriminative power while obtaining a system ranking list similar to the canonical version at the same time.
\end{abstract}

\ccsdesc[500]{Information systems~Retrieval effectiveness}

\keywords{offline evaluation, evaluation metrics, meta-evaluation, discriminative power}

\maketitle

\section{Introduction}
Online and offline evaluations of ranked retrieval systems complement each other to advance the state of the art of web search engines and other ranking applications. Recently, Moffat \textit{et al.}~\cite{Moffat-2022} proposed an analytic framework, namely C/W/L/A, for offline evaluation metrics. Under the framework, the score of a metric can be obtained from the combination of a user browsing model given by the  \textit{continuation probability} ($C(\cdot)$) and a user gain \textit{aggregation} function ($A(\cdot)$) as the following. 
\begin{equation}
    \label{cwla}
    M_\mathrm{CWLA}(\mathbf{r}) = \sum_{i=1}^{\infty}{L(i)\cdot A(i)}
\end{equation}
Here, $M_\mathrm{CWLA}$ is the metric score, $\mathbf{r} = <r_1, r_2, \cdots, r_i>$ is the relevance levels of the documents from position $1$ to $i$, $A(i)$ represents how users accumulate their gain when they end the search interaction at rank $i$, and $L(i)$ is the probability a user inspects the item at position $i$ and then stop inspecting the search engine result page (SERP). $L(i)$ can be eventually obtained from $C(i)$ , the probability that a user who has inspected the $i$-th item in the SERP will continue to examine the item at rank $i+1$, through:

\begin{equation}
    \label{l-c}
    L(i) = (1 - C(i))\prod_{j=1}^{i-1}C(j)
\end{equation}

C/W/L/A framework provides the information retrieval (IR) community with numerous alternative evaluation metrics through the flexible combination of $C(\cdot)$ and $A(\cdot)$. For example, one can combine the browsing model of Discounted Cumulative Gain~(DCG)~\cite{Jarvelin-2002}~(refer to Eq.\ref{eq:dcg}) and an aggregation function $A_\text{PE}$ assuming that the user’s gain from the SERP is in compliance with \textit{the peak-end rule}~\cite{Fredrickson-1993-Duration}~(refer to Eq.\ref{eq:pe}), to create a new evaluation metric.

These metrics have the potential to be used for evaluating IR systems offline from multiple different perspectives. However, the statistical stability of these alternative evaluation metrics in offline evaluation is not yet investigated. In particular, in order for alternative metrics obtained by combining various browsing models and different aggregations to be widely used in offline evaluations, the IR community needs to understand that, given a browsing model, how metrics obtained from different aggregations perform in terms of statistical stability. 

Inspired by previous work~\cite{Sakai-2021-ecir}, in this study, we investigate the statistical stability of C/W/L/A metrics in offline evaluation. Specifically, we come up with the following \textbf{research questions} (RQs).

\begin{itemize}
\item \textbf{RQ1}:~Given a browsing model, how metrics obtained from different aggregations resemble one another?~(The system ranking similarity among aggregations)
\item \textbf{RQ2}:~Given a browsing model, how metrics obtained from different aggregations perform in terms of system ranking consistency across two disjoint topic sets?~(The system ranking consistency of aggregations)
\item \textbf{RQ3}:~Given a browsing model, how metrics obtained from different aggregations perform in terms of discriminative power, the ability to tell one run is better than another with statistical significance?~(The discriminative power of aggregations)

\item Moreover, after investigating the above three questions, we further propose \textbf{RQ4}: Can we find an alternative metric(s) that improves system ranking consistency or (and) discriminative power compared to the canonical version, while at the same time returning system ranking lists that are very similar to the canonical version?
\end{itemize}

To figure out these research questions, we combined various aggregation functions with the browsing model of Precision, Discounted Cumulative Gain~(DCG)~\cite{Jarvelin-2002}, Rank-Biased Precision~(RBP)~\cite{Moffat-2008}, INST~\cite{Bailey-2015}, Average Precision (AP)~\cite{Sanderson-2005} and Expected Reciprocal Rank~(ERR)~\cite{Chapelle-2009}~(Section ~\ref{sec:metric}); and then we conducted experiments on two offline evaluation collections, examining performances of aggregations in terms of system ranking similarity~(Section~\ref{sec:sim}), system ranking consistency~(Section~\ref{sec:const}) and discriminative power~( \ref{sec:disc}). 

Our result suggests that, in terms of system ranking consistency and discriminative power, the aggregation function of \textit{expected rate of gain} ($A_{\text{ERG}}$) has an outstanding performance while the aggregation function of \textit{maximum relevance} ($A_{\text{max}}$) usually has an insufficient performance. Our result also suggests that Precision, DCG, RBP, INST and AP with their canonical aggregation all have favourable performances in system ranking consistency and discriminative power. But for ERR, replacing its canonical aggregation with ERG can further strengthen the discriminative power while obtaining a system ranking list similar to the canonical version at the same time. 
 
As far as we know, we are the first to examine the statistical stability of metrics generated by C/W/L/A framework. Our work extends the work of Moffat \textit{et al}~\cite{Moffat-2022} from the perspective of statistical reliability in offline evaluation experiment. Based on the result, we suggest researchers who want to design reliable evaluation metrics with C/W/L/A framework use ERG as the aggregation function in order to achieve favourable system ranking consistency and discriminative power.

\section{Related work}
Evaluating the effectiveness of search engines has long been a central concern for the information retrieval~(IR) community. Existing evaluation methods can be broadly divided into two classes, online evaluation and  offline evaluation~\cite{Voorhees-2002}. Offline evaluation is often built upon different simulations of the process of a user interacting with a system under operational settings~\cite{Sanderson-2010}, and the evaluation metric scores can be viewed as the simulation of the \textit{gain} a user accumulated during that process. Widely used offline evaluation metrics include: Precision, DCG~\cite{Jarvelin-2002}, AP~\cite{Sanderson-2005},  RBP~\cite{Moffat-2008}, ERR~\cite{Chapelle-2009}, INST~\cite{Bailey-2015},~among others~\cite{Moffat-2012, Zhang-2017, Azzopardi-2018-IFT, Azzopardi-2021-ERR}.

\subsection{The C/W/L/A Framework}
To characterise user models behind offline evaluation metrics, C/W/L framework~\cite{Moffat-2013-Users, Moffat-2017-Incorporating} was proposed. In C/W/L framework, the user model behind a metric can be deconstructed into three interrelated aspects of the user behavior:
\begin{itemize}
    \item \textit{Continuation probability},~$C(i)$:~the probability that a user who has inspected the $i$-th item in the SERP will continue to examine the item at rank $i+1$.
    \item \textit{Weight function},~$W(i)$:~ the fraction of user attention on the item at position $i$. In other words, it is the likelihood of a user viewing the item at position $i$ at any time under a sequence of random selections. 
    \item \textit{Last probability},~$L(i)$:~the probability that a user examine the document at rank $i$ and then stop interacting with the SERP. 
\end{itemize}

In the C/W/L framework, as long as one of the three components is known, the other two components can be calculated as well. For example, the $L(i)$ can be calculated by $C(i)$ through Eq.~\ref{l-c}.

The C/W/L framework thus provides a common ground for comparing models of various widely used metrics like DCG and RBP in terms of user browsing behaviour. For instance, the user model of Rank-Biased Precision~(RBP)@$p$ can be viewed as the assumption that at each rank of the SERP, the user will continue to examine the next result with a constant probability of $p$~\cite{Moffat-2013-Users}. The C/W/L framework also allow researchers to design new metrics by defining $C(i)$~\cite{Azzopardi-2018-IFT, azzopardi-2020-data, Azzopardi-2021-ERR}. 

However, the original C/W/L framework assumes that users accumulate their gain only through the form of \textit{Expected Rate of Gain} (also known as \textit{Expected Utility}, refer to Eq.~\ref{eq:erg}) or \textit{Expected Total Gain} (refer to Eq.~\ref{eq:etg}). This assumption cannot explain the user behavior behind ERR, a metric widely used in offline evaluation practice, since there is no appropriate aggregation for ERR given its browsing model~(refer to Eq.~\ref{eq:err})~\cite{Azzopardi-2021-ERR}. To resolve this incompatibility limitation, Moffat \textit{et al.}~\cite{Moffat-2022} extended the C/W/L framework to the C/W/L/A framework by introducing a new component: \textit{aggregation}~(A). The score of a metric under the C/W/L/A framework can be computed through Eq.~\ref{cwla}. 

The introduction of \textit{aggregation} allowed the C/W/L/A framework to characterize metrics like ERR through incorporating an appropriate aggregation function. For example, the score of ERR can be computed from the combination of $C_{\text{ERR}}$~(Eq.~\ref{eq:err}) and $A_{\text{ERR}}$~(Eq.~\ref{eq:aerr}) .

\subsection{Meta-Evaluation of Metrics}
As various evaluation metrics have been proposed, IR researchers have to come to grips with a question: what should a “good” evaluation metric be? Driven by the question, researchers began to shed light on the meta-evaluation of evaluation metrics. Sakai~\cite{Sakai-2021-ecir} argued that, as offline evaluation measures are used in experiments in the hope of ameliorating the effectiveness of search systems for real users, a good evaluation metric should:~(a) serve as surrogates of users’ perspectives so that IR systems can be improved align with the better user experience~(user satisfaction); and~(b)~be statistically stable so that reliable offline experiments can be conducted~(statistical stability). 

As \textit{user satisfaction} is regarded as a near-ideal ground truth metric of retrieval effectiveness, meta-evaluating metrics from the perspective of user satisfaction has already been widely adopted in previous studies~\cite{maskari-2008, sanderson-2010-preference}. To measure to what extent metric scores are consistent with users' satisfaction feedbacks, some researchers use correlations with users' satisfaction feedbacks~\cite{Chen-2017-THUIR1, Zhang-2020-Models, Wicaksono-2020-Metrics, Liu-2021-CIKM}, while others use agreements with users' SERP preference~\cite{sanderson-2010-preference}. 
Moffat \textit{et al.}~\cite{Moffat-2022} have already meta-evaluated the performances of aggregation functions in terms of the correlation with users' satisfaction feedbacks and the agreement with users' SERP preferences. Experimental results from Moffat \textit{et al.}~\cite{Moffat-2022} showed that the aggregation function using \textit{maximum relevance} ($A_{\text{max}}$) usually correlates well with users' satisfaction feedbacks, but in terms of agreement users' SERP preferences,  the aggregation function using \textit{expected rate of gain} (ERG) usually performs better. However, Moffat \textit{et al.}~\cite{Moffat-2022} did not meta-evaluate the performances of aggregation functions in terms of the statistical stability. Hence the present study complements their work.

\textit{Discriminative power}~\cite{Sakai-2006}, the statistical ability of a metric to significantly discriminate system pairs, is a widely used method to meta-evaluate the statistical stability of metrics~\cite{Anelli-2019, Kanoulas-2009, robertson-2010}. Discriminative power measures the \textit{stability} of a metric across the topics based on significance testing~\cite{Sakai-2014-promise}. 

\textit{System ranking consistency}, which is based on swap method is another method to meta-evaluate the statistical stability of metrics~\cite{Zobel-1998,  Voorhees-1998, Buckley-2000, Voorhees-2002-effect, Voorhees-2009}. System ranking consistency is the similarity of two rankings given by an evaluation measure on topic set A and topic set B respectively~\cite{Sakai-2021-ecir, amigo-etal-2020-effectiveness}. Previous work~\cite{Sakai-2021-ecir} formalised the procedure to measure system ranking consistency as randomly spliting the topic set multiple times and completing with distribution-free statistical significance testing for the difference in mean $\tau$’s between two topic subsets. 

Other meta-evaluation methods include \textit{judgement cost}~\cite{Bttcher-2007}, \textit{coverage}~\cite{ravana-2010}, and axiomatic approaches~\cite{Amigo-2018}, but they are beyond the scope of this study. 

In this study, we focus on the statistical stability of various aggregation functions in the C/W/L/A framework. In contrast to previous work (\textit{(e.g.,}~\cite{Sakai-2021-ecir}), we are concerned with \textit{the within-group difference of metrics under the same browsing model while adopting different aggregations} rather than the between-group difference of metrics under different browsing models. Note that the statistical stability of a metric cannot tell whether the metrics is``measuring what we want to measure''~\cite{Sakai-2014-promise}~(e.g., how well a metric is correlating with users' satisfaction feedback). Thus it meta-evaluates metrics on a dimension orthogonal to user satisfaction. 

\section{Experimental Settings}
\begin{table*}
  \caption{Combinations of $C(\cdot)$ and aggregations examined in the experiment. The asterisk indicates the canonical aggregation of each metric. An {\texttimes} mark indicates a combination where the metric score is a constant and thus is impractical.}
  \label{tab:cmb}
  \begin{tabular}{lccccccc}
    \toprule
      & $A_{\text{ERG}}$ & $A_{\text{ETG}}$  & $A_{\text{avg}}$ & $A_{\text{max}}$ & $A_{\text{fin}}$ & $A_{\text{PE}}$ & $A_{\text{ERR}}$\\
    \midrule
    $C_{\text{Precision}@10}$ &  Precision\_ERG * & Precision\_ETG & Precision\_avg &  Precision\_max & Precision\_fin &  Precision\_PE & {\texttimes} \\
    $C_{\text{DCG}@10}$ &  DCG\_ERG*  & DCG\_ETG & DCG\_avg &  DCG\_max & DCG\_fin &  DCG\_PE & {\texttimes} \\
    $C_{\text{RBP}@p=0.8}$ & RBP\_ERG* & RBP\_ETG & RBP\_avg &  RBP\_max & RBP\_fin &  RBP\_PE & {\texttimes} \\
    $C_{\text{INST}@T=2.5}$ &  INST\_ERG  *  & INST\_ETG& INST\_avg &  INST\_max & INST\_fin &  INST\_PE & INST\_ERR \\
    $C_{\text{AP}}$ &  AP\_ERG* & AP\_ETG & AP\_avg &  AP\_max & AP\_fin &  AP\_PE & AP\_ERR \\
    $C_{\text{ERR}}$ & ERR\_ERG  & ERR\_ETG & ERR\_avg &  ERR\_max & ERR\_fin &  ERR\_PE & ERR\_ERR* \\
  \bottomrule
\end{tabular}
\end{table*}

\subsection{Metrics}
\label{sec:metric}
In our experiment, we consider browsing models of Precision$@k=10$, DCG$@k=10$,  RBP$@p=0.8$, INST$@T=2.25$, AP and ERR. These metrics and parameters are chosen because:~(1)~they have clearly defined browsing models in C/W/L/A framework;~(2)~combinations of their browsing models and different aggregations have already examined by Moffat \textit{et al.}~\cite{Moffat-2022} in terms of consistency with user satisfaction. The browsing models of these metrics in C/W/L/A framework are as follows. 
\begin{itemize}
    \item Precision$@k$: 
    \begin{equation}
    \label{eq:prec}
    C_{\text{Precision}}(i)  = 
    \left\{
        \begin{aligned}
        & 1  & \mathrm{for\quad} i < k, \\
        & 0 & \mathrm{otherwise.}
        \end{aligned}
    \right.
    \end{equation}
    
    \item Discounted Cumulative Gain at $k$ (DCG$@k$)~\cite{Jarvelin-2002}:
    \begin{equation}
    \label{eq:dcg}
    C_{\text{DCG}@k}(i) = 
    \left\{
        \begin{aligned}
        & \frac{\log_2(i+1)}{\log_2(i+2)}  & \mathrm{for\quad} i < k, \\
        & 0 & \mathrm{otherwise.}
        \end{aligned}
    \right.
    \end{equation}
    
    \item Rank-Biased Precision (RBP$@p$)~\cite{Moffat-2008}:
    \begin{equation}
    \label{eq:rbp}
    C_{\text{RBP}@p}(i) = p
    \end{equation}   

    \item INST$@T$~\cite{Bailey-2015}:
    \begin{equation}
    \label{eq:inst}
    C_{\text{INST}@T}(i) = \frac{(i - 1 + T + T_i)^2}{(i + T + T_i)^2} 
    \end{equation} 
    where $T_i =T - \sum_{j = 1}^{i} r_i$ represents the remaining gain the user needs to acquire in order to fulfill the expected gain after inspecting the $i$-th item.
    
    \item Average Precision (AP)~\cite{Sanderson-2005}:
    \begin{equation}
    \label{eq:ap}
    C_{\text{AP}}(i) = \frac{\sum_{j = i+1}^{\infty} (r_j / j)}{\sum_{j = i}^{\infty} (r_j / j)}
    \end{equation}  
    
    \item Exponential Reciprocal Rank (ERR)~\cite{Chapelle-2009}:
    \begin{equation}
    \label{eq:err}
    C_{\text{ERR}}(i) = 1 - r_i
    \end{equation}
\end{itemize}

In our experiment, we combine the $C(\cdot)$ of the above metrics with the following \textit{aggregation} functions which can be referred to the work of Moffat \text{et al.}~\cite{Moffat-2022}. Table \ref{tab:cmb} shows the combinations examined in our experiment. 

\begin{itemize}
    \item The \textit{expected rate of gain}~($A_\text{ERG}$):
    \begin{equation}
        \label{eq:erg}
        A_{\text{ERG}}(i) = \frac{1}{V^+}\sum_{j = 1}^{i} r_i
    \end{equation}
    where
    \begin{equation*}
         \frac{1}{V^+} = \sum_{i = 1}^{\infty}\prod_{j = 1}^{i-1}C(j)
    \end{equation*}    
    
    This aggregation represents the ``expected utility accumulated per item inspected'' in the original C/W/L framework~\cite{Moffat-2013-Users, Moffat-2017-Incorporating}. 
    
    \item The \textit{expected total gain}~($A_\text{ETG}$):
    \begin{equation}
        \label{eq:etg}
        A_{\text{ETG}}(i) = \sum_{j = 1}^{i} r_i
    \end{equation}
    
    This aggregation function assumes that users simply sum up the gain collected from each item with the same weight when they leave.
     
    \item The average relevance~($A_\text{avg}$):
    \begin{equation}
        \label{eq:avg}
        A_{\text{avg}}(i) = \frac{1}{i}\sum_{j = 1}^{i} r_i
    \end{equation}
    
    This aggregation function assumes that users' gain is determined by the average relevance of items they inspected  when they leave.
    
    \item The maximum relevance ($A_\text{max}$)
    \begin{equation}
        A_{\text{max}}(i) = {\max}_{j = 1}^{i} r_j
    \end{equation}

    This aggregation function assuming that the user’s gain from the SERP will be completely dominated by the \emph{best} element they inspected when they leave:
    
    \item The last relevance ($A_\text{fin}$)
    \begin{equation}
        A_{\text{fin}}(i) = r_i
    \end{equation}
    
    This aggregation function assuming that the user’s gain from the SERP will be completely dominated by the \emph{last} element they observed when they leave.

    \item  The peak-end relevance ($A_\text{PE}$)
    \begin{equation}
        \label{eq:pe}
        A_{\text{PE}}(i) = \beta \cdot A_{\text{max}}(i) + (1-\beta)\cdot A_{\text{fin}}(i)
    \end{equation}
    
    This aggregation function assuming that the user’s gain from the SERP is in compliance with \textit{the peak-end rule}, which suggests that people judge their experience of a series of past events by how they felt at its peak and by what they occurred most recently~\cite{Fredrickson-1993-Duration}. 
    In our experiment we set $\beta$ = 0.5 following the setting in the work of Moffat \textit{et al.}~\cite{Moffat-2022}.

    \item  The aggregation for ERR ($A_\text{ERR}$)
    \begin{equation}
        \label{eq:aerr}
        A_{\text{ERR}}(i) = 1/i
    \end{equation}
    
    This aggregation function assuming that the user becomes increasingly dissatisfied as he or she inspects more documents, regardless the quality of the document. 
\end{itemize}

\subsection{Dataset}
\begin{table}
  \caption{Overview of datasets used in our experiment.}
    \label{tab:dataset}
   \begin{tabular}{cccccc}
        \toprule
        Our short Name&\#topics & rel. levels & \#rel. per topic & \#runs\\
        \midrule
        WWW3 & 80 & 4 & 159.0 & 39\\
        TR19DL & 43 & 4 & 153.4 & 38\\
        \bottomrule
   \end{tabular}
\end{table}

Table~\ref{tab:dataset} presents an overview of datasets we used in our experiment, where we examine the system ranking similarity, system ranking consistency and discriminative power of the metrics listed in Table~\ref{tab:cmb}. These datasets were chosen based on the following principles: (1)~they should be recent;~(2)~they should include enough topics and submitted runs, since we want to obtain reliable experimental results.

The NTCIR-15 WWW-3 (WWW3)~\cite{Sakai-2021-WWW3} dataset is from the NTCIR-15 WWW-3 English subtask whose target corpus is clueweb12-B13 (about 50 million web pages)~\footnote{https://lemurproject.org/clueweb12/}. It includes 80 topics and 39 runs~(including 2 baseline runs), with 4-level relevance judgement for documents. 

The target corpus for the TREC 2019 Deep Learning track (TR19DL) dataset is an MS MARCO corpus (3.2 million documents). TR19DL dataset includes 43 topics and 38 runs, with 4-level relevance judgement for documents. 

When calculating scores for Precision, DCG, RBP, INST and AP, we linearly map the relevance score of the $i$-th item in the form of $r_i = x/x_{max}$. When calculating scores for ERR, we exponentially the relevance score of the $i$-th item in the form of $r_i = (2^{x} - 1)/2^{x_{max}}$. Here $x$ is the original relevance score, ${x_{max}}$ is the maximum relevance score in the collection. We exponentially map relevance scores for ERR in order to make its metric scores be identical to ones given by its original definition~\cite{Chapelle-2009}.

\section{System Ranking Similarity}
\label{sec:sim}

\begin{figure*}[htbp]
\centering
\begin{subfigure}[t]{0.33\textwidth}
    \centering
    \includegraphics[width=5.5cm]{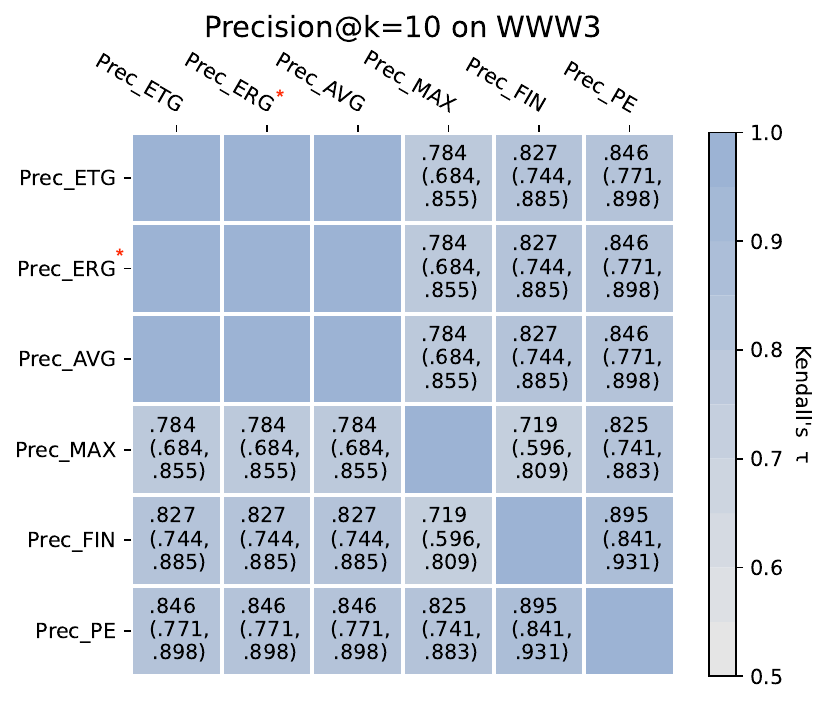}
    \caption*{}
\end{subfigure}
\begin{subfigure}[t]{0.33\textwidth}
    \centering
    \includegraphics[width=5.5cm]{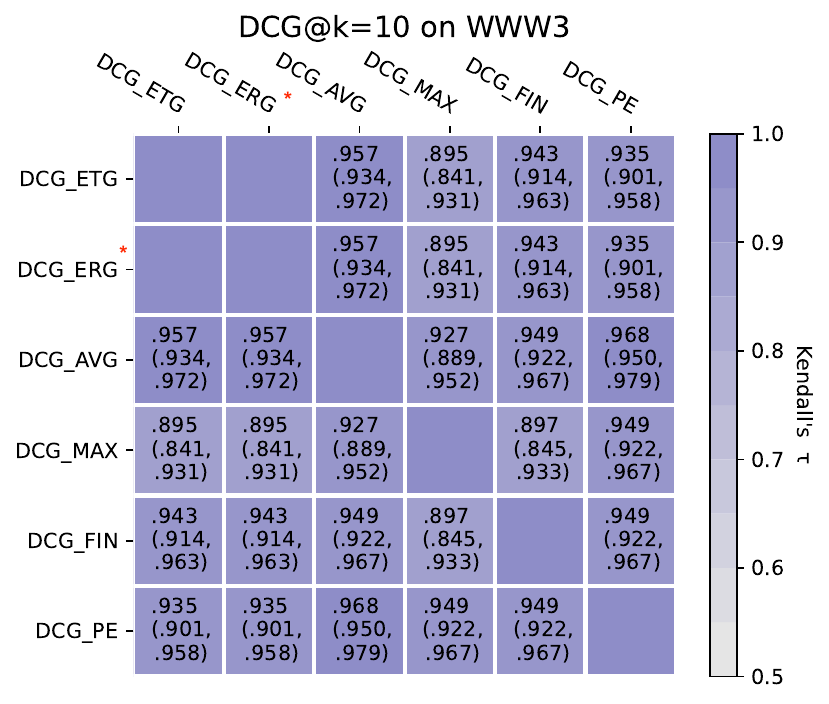}
    \caption*{}
\end{subfigure}
\begin{subfigure}[t]{0.33\textwidth}
    \centering
    \includegraphics[width=5.5cm]{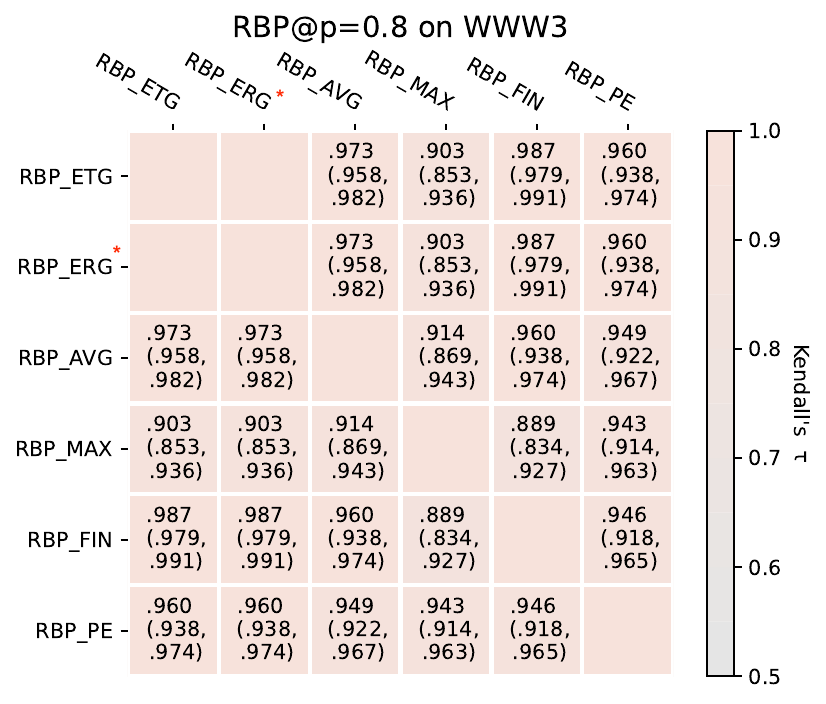}
    \caption*{}
\end{subfigure}
\begin{subfigure}[t]{0.33\textwidth}
    \centering
    \includegraphics[width=5.5cm]{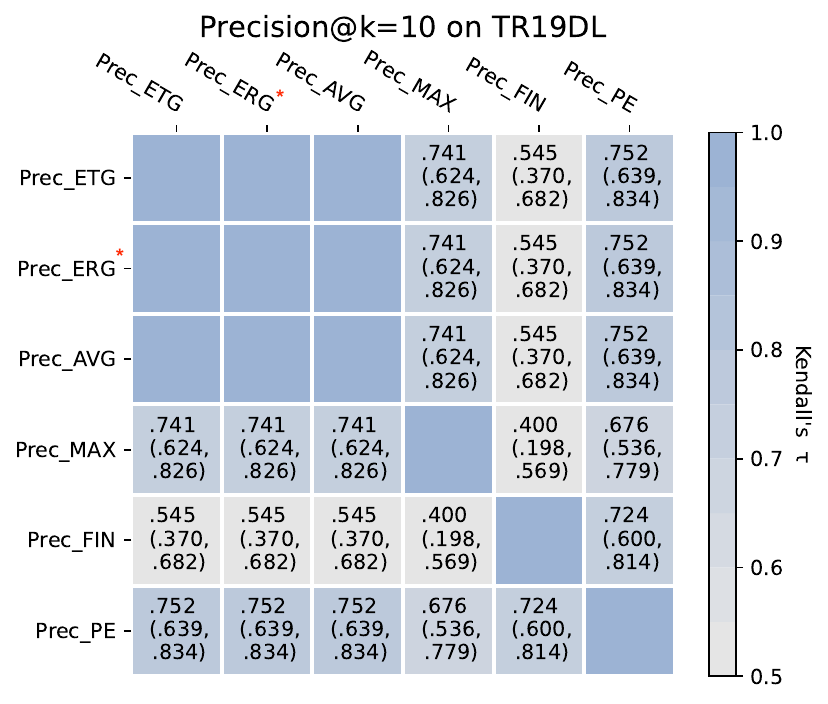}
    \caption*{}
\end{subfigure}
\begin{subfigure}[t]{0.33\textwidth}
    \centering
    \includegraphics[width=5.5cm]{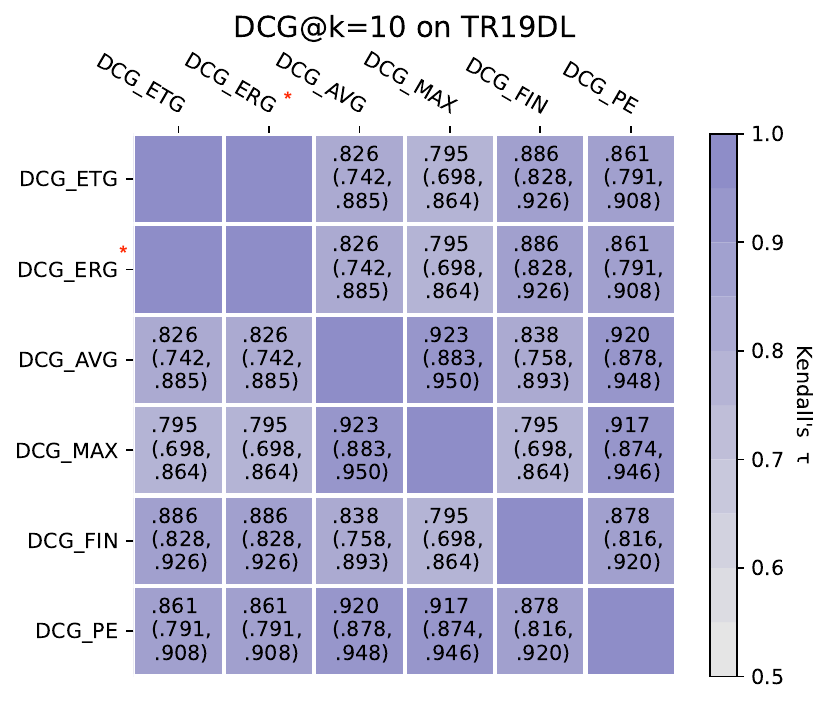}
    \caption*{}
\end{subfigure}
\begin{subfigure}[t]{0.33\textwidth}
    \centering
    \includegraphics[width=5.5cm]{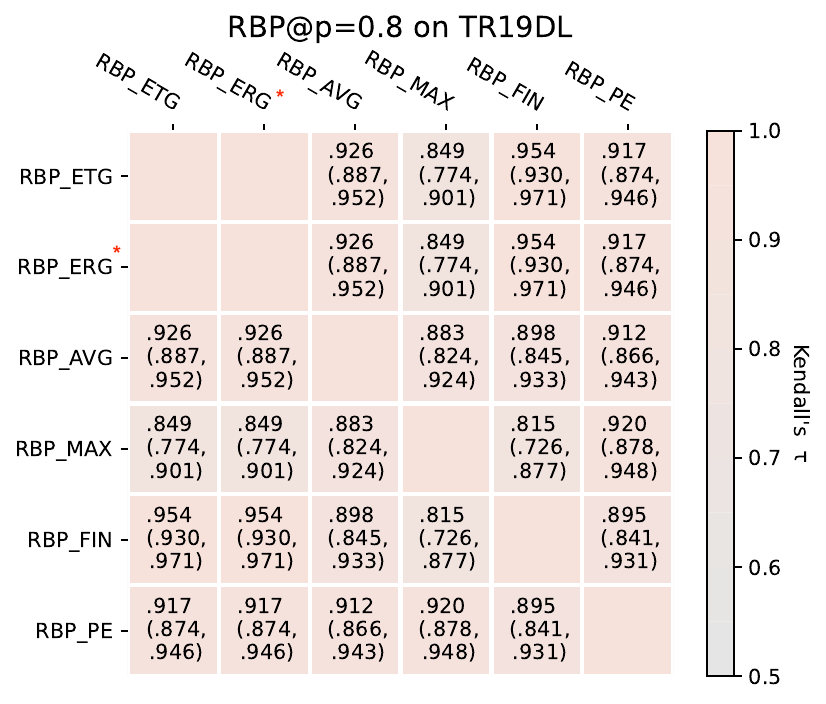}
    \caption*{}
\end{subfigure}
\begin{subfigure}[t]{0.33\textwidth}
    \centering
    \includegraphics[width=5.5cm]{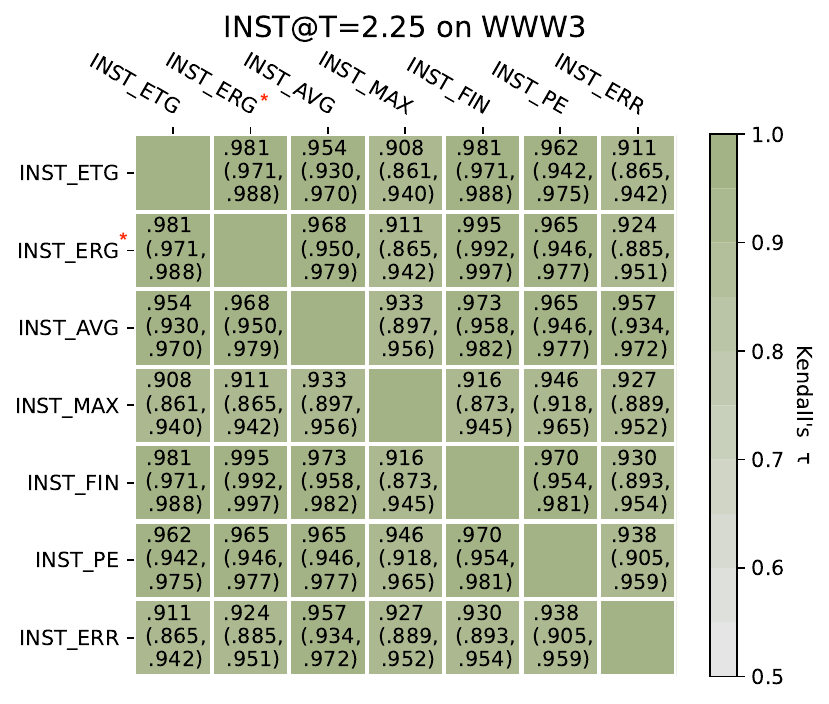}
    \caption*{}
\end{subfigure}
\begin{subfigure}[t]{0.33\textwidth}
    \centering
    \includegraphics[width=5.5cm]{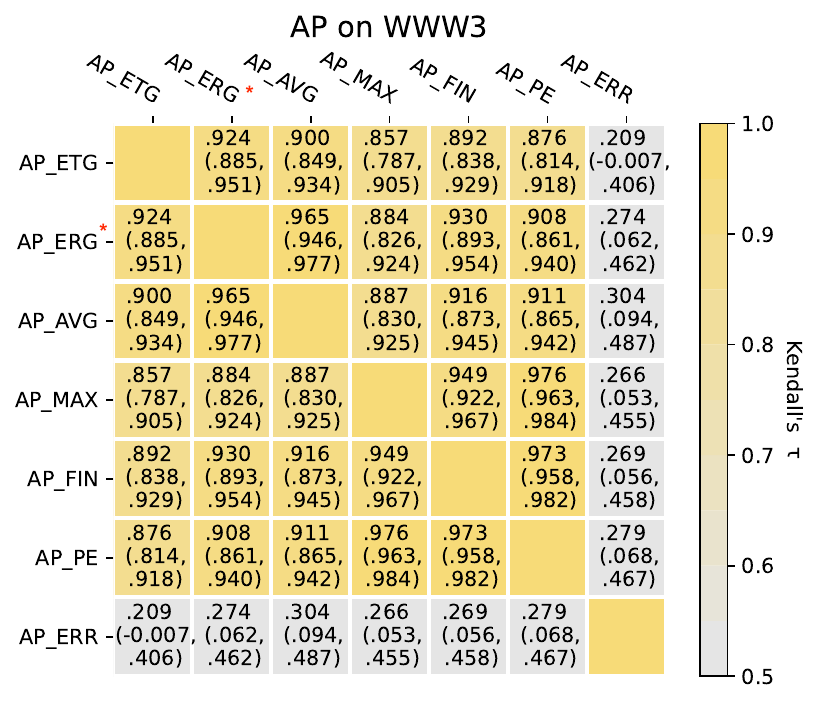}
    \caption*{}
\end{subfigure}
\begin{subfigure}[t]{0.33\textwidth}
    \centering
    \includegraphics[width=5.5cm]{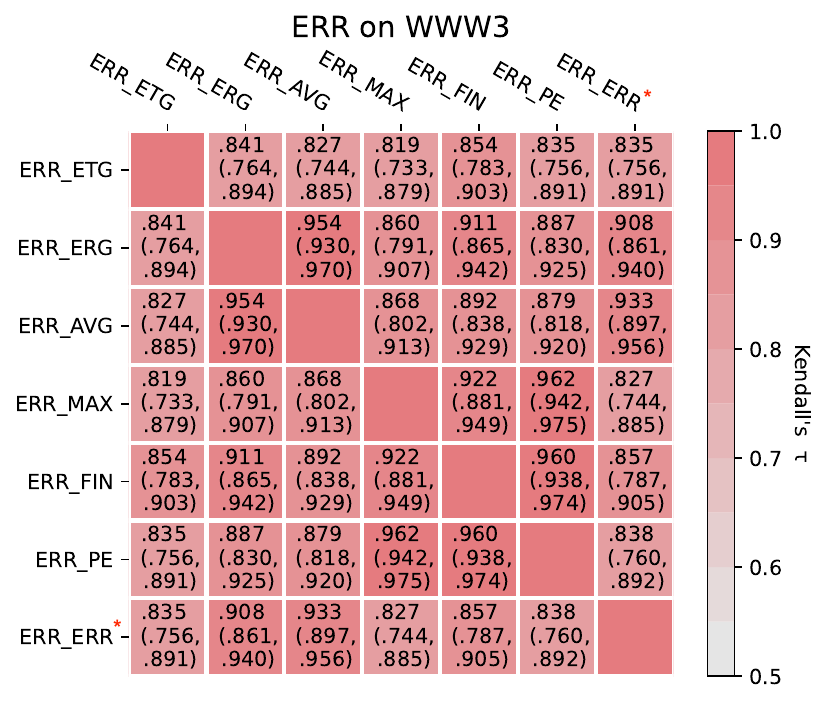}
    \caption*{}
\end{subfigure}
\end{figure*}
\clearpage

\begin{figure*}[htbp]\ContinuedFloat
\centering
\begin{subfigure}[t]{0.33\textwidth}
    \centering
    \includegraphics[width=5.5cm]{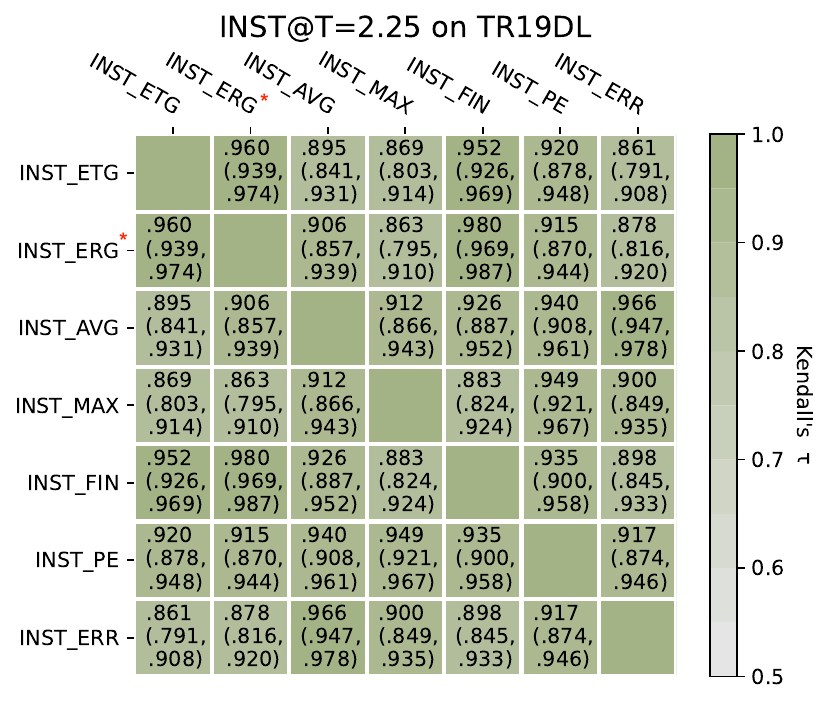}
    \caption*{}
\end{subfigure}
\begin{subfigure}[t]{0.33\textwidth}
    \centering
    \includegraphics[width=5.5cm]{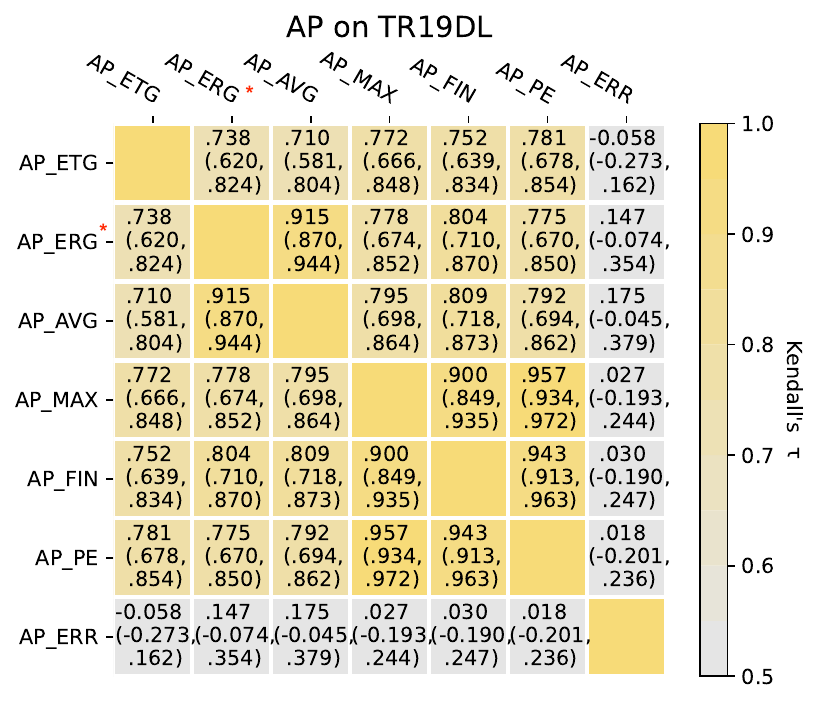}
    \caption*{}
\end{subfigure}
\begin{subfigure}[t]{0.33\textwidth}
    \centering
    \includegraphics[width=5.5cm]{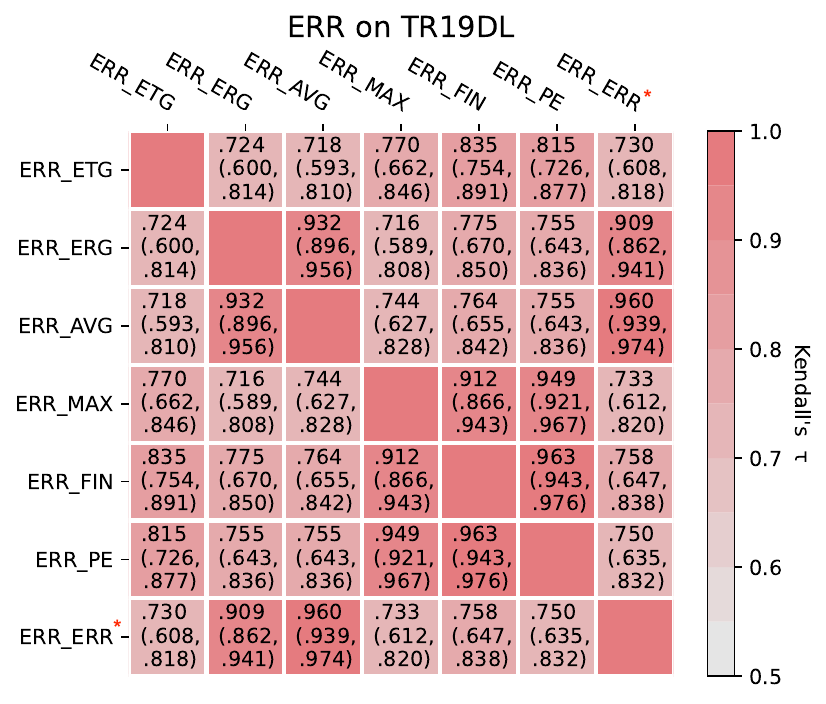}
    \caption*{}
\end{subfigure}
\caption{Ranking similarity in terms of Kendall's $tau$ of C/W/L/A Metrics on the two datasets. The asterisk indicates the canonical aggregation of each metric.}
\Description{Ranking similarity in terms of Kendall's $tau$ of C/W/L/A Metrics on the two datasets. The asterisk indicates the canonical aggregation of each metric.}
\label{fig:corr}
\end{figure*}

Figure \ref{fig:corr} shows how the system rankings according to different aggregations under the browsing model of a metric resemble one another in terms of Kendall's $\tau$. 95\% CIs for correlations are given in parentheses.

Note that for Precision, DCG and RBP, the system ranking list returned by $A_{\text{ERG}}$ and $A_{\text{ETG}}$ are exactly the same. That is because for these metrics, the $C(i)$ is a constant given $i$~(refer to Eq.~\ref{eq:prec}, Eq.~\ref{eq:dcg} and Eq.~\ref{eq:rbp}), the metric score given by $A_{\text{ETG}}$ is thus equal to the metric score given by $A_{\text{ERG}}$ multiplying with a constant~(refer to Eq.~\ref{eq:erg} and Eq.~\ref{eq:etg}). For Precision, the system ranking list returned by $A_{\text{avg}}$ is the same as ones returned by $A_{\text{ERG}}$ and $A_{\text{ETG}}$. That is because for Precision$@k$, $L(i) = 1$ when $i = k$ and $L(i) = 0$ while in other cases, the metric score of Precision$@k$ is thus equal to $A(k)$~(refer to Eq.~\ref{cwla}, Eq.~\ref{l-c} and Eq.~\ref{eq:prec}). Hence, when $k$ is given, the metric score of Precision$@k$ given by $A_{\text{avg}}$ is thus equal to the one given by $A_{\text{ETG}}$ multiplying with a constant~(refer to Eq.~\ref{eq:etg} and Eq.~\ref{eq:avg}).

From Figure \ref{fig:corr} we can observe that:~(1)~the system ranking similarity among different aggregations depends on the browsing model of a metric and it is hard, if not impossible, to summarise in a few words. More specifically, the following can be observed in terms of system ranking similarity;~(2)~generally, system ranking lists returned by different aggregations under the browsing model of a metric are more similar on WWW3 compared to results on TR19DL. We can further observe the following result.

The system ranking similarity among different aggregations tends to be low under the browsing model of Precision. No pair of aggregations has a system ranking similarity of more than $0.90$ in terms of $\tau$ on WWW3 dataset, and the $\tau$ is even lower on TR19DL dataset.

Under the browsing model of DCG, system rankings given by different aggregations tends to be similar to each other on WWW3, while on TR19DL, system ranking lists returned by $ A_{\text{max}}$ tend to be less similar to system ranking lists returned by $ A_{\text{ETG}}$, $ A_{\text{ERG}}$ and $ A_{\text{fin}}$.

Under the browsing model of RBP, system rankings given by different aggregations tends to be similar each other on both of the two datasets. Specifically, system ranking lists returned by $ A_{\text{max}}$ are relatively less similar to system ranking lists returned by $ A_{\text{ETG}}$, $ A_{\text{ERG}}$ and $ A_{\text{fin}}$. 

Under the browsing model of INST, system rankings given by different aggregations also tends to be similar each other on both of the two datasets. Specifically, system ranking lists returned by $ A_{\text{max}}$ are relatively less similar to system ranking lists returned by $ A_{\text{ETG}}$ and $ A_{\text{ERG}}$.

 Under the browsing model of AP, it is clear that:~(1)~System ranking lists returned by $ A_{\text{ERR}}$ is very different to ones returned by other aggregations;~(2)~$ A_{\text{avg}}$ has high system ranking similarity to $ A_{\text{ERG}}$;~(3)~system ranking lists returned by $ A_{\text{max}}$, $ A_{\text{fin}}$ and $ A_{\text{PE}}$ are similar to each other.

 Under the browsing model of ERR, it is clear that:~(1)~System ranking lists returned by $ A_{\text{ETG}}$ is less similar to ones returned by other aggregations;~(2)~system ranking lists returned by $ A_{\text{ERG}}$, $ A_{\text{avg}}$ and $ A_{\text{ERR}}$ are similar to each other;~(3)~system ranking lists returned by $ A_{\text{max}}$, $ A_{\text{fin}}$ and $ A_{\text{PE}}$ are similar to each other.
\section{System Ranking Consistency}
\label{sec:const}

\begin{table*}
  \caption{System ranking consistency of metrics with different aggregations in terms of mean $\tau$ over $B$ = 1,000 trials on the two data sets. The asterisk indicates the canonical aggregation of each metric. \EightFlowerPetal/\SixFlowerPetalDotted/\SixFlowerAlternate/\FiveFlowerOpen/\FourClowerOpen/\OrnamentDiamondSolid~means statistically significantly ($p < 0.05$) outperforms the worst 6/5/4/3/2/1 aggregation(s) respectively.}
  \label{tab:const}
  \begin{tabular}{llllllllllll}
    \toprule
         \multicolumn{12}{c}{WWW3 (80 topics)}\\
    \midrule
     \multicolumn{2}{c}{$C_{\text{Precision}@10}$}& \multicolumn{2}{c}{$C_{\text{DCG}@10}$} & \multicolumn{2}{c}{$C_{\text{RBP}@p=0.8}$} &  \multicolumn{2}{c}{$C_{\text{INST}@T=2.5}$} & \multicolumn{2}{c}{$C_{\text{AP}}$}  &  \multicolumn{2}{c}{$C_{\text{ERR}}$}\\
    \midrule
      $ A_{\text{ERG}}$* \FiveFlowerOpen & 0.755 & $ A_{\text{ERG}}$* \SixFlowerAlternate & 0.763  &  $ A_{\text{fin}}$ \FiveFlowerOpen & 0.765 &   $ A_{\text{ERG}}$* \SixFlowerPetalDotted & 0.760 & $ A_{\text{ETG}}$ \SixFlowerAlternate & 0.756 & $ A_{\text{ERG}}$ \EightFlowerPetal & 0.723 \\
      
      $ A_{\text{ETG}}$ \FiveFlowerOpen & 0.755 & $ A_{\text{ETG}}$ \SixFlowerAlternate & 0.763  & $ A_{\text{ERG}}$* \FiveFlowerOpen & 0.762 &  $ A_{\text{fin}}$ \SixFlowerAlternate & 0.757  &  $ A_{\text{ERG}}$* \SixFlowerAlternate & 0.753 & $ A_{\text{ERR}}$ * \SixFlowerAlternate & 0.701\\
      
      $ A_{\text{avg}}$ \FiveFlowerOpen& 0.755 & $ A_{\text{fin}}$ \FiveFlowerOpen& 0.744 &  $ A_{\text{ETG}}$ \FiveFlowerOpen & 0.762 &   
      $ A_{\text{ETG}}$  \SixFlowerAlternate & 0.753 &  $ A_{\text{avg}}$ \FourClowerOpen & 0.721 & $ A_{\text{avg}}$ \SixFlowerAlternate & 0.701\\
      
     $ A_{\text{PE}}$ \FourClowerOpen & 0.641 & $ A_{\text{avg}}$ \OrnamentDiamondSolid & 0.735 & $ A_{\text{PE}}$ \FourClowerOpen & 0.745 &  
     $ A_{\text{PE}}$ \FiveFlowerOpen & 0.732 & $ A_{\text{fin}}$ \OrnamentDiamondSolid & 0.706 & $ A_{\text{ETG}}$ \FourClowerOpen & 0.672\\
     
     $ A_{\text{max}}$ \OrnamentDiamondSolid & 0.600 & $ A_{\text{PE}}$ \OrnamentDiamondSolid & 0.734  &  $ A_{\text{avg}}$ \OrnamentDiamondSolid & 0.740 & 
     $ A_{\text{avg}}$ \FourClowerOpen & 0.723 & $ A_{\text{PE}}$ \OrnamentDiamondSolid & 0.690 & $ A_{\text{fin}}$ \FourClowerOpen & 0.666\\
     
     $ A_{\text{fin}}$ & 0.552 & $ A_{\text{max}}$ & 0.687  &  $ A_{\text{max}}$ & 0.681 & 
     $ A_{\text{ERR}}$ \OrnamentDiamondSolid & 0.692 & 
     $ A_{\text{max}}$ \OrnamentDiamondSolid& 0.668 & $ A_{\text{PE}}$ \OrnamentDiamondSolid& 0.651\\
     
     &   &   &     &   &   & $ A_{\text{max}}$ & 0.682 &
     $ A_{\text{ERR}}$ & -0.090 & $ A_{\text{max}}$ & 0.619\\
    \bottomrule
    \toprule
    \multicolumn{12}{c}{TR19DL (43 topics)}\\
    \midrule
     \multicolumn{2}{c}{$C_{\text{Precision}@10}$}& \multicolumn{2}{c}{$C_{\text{DCG}@10}$} & \multicolumn{2}{c}{$C_{\text{RBP}@p=0.8}$} &  \multicolumn{2}{c}{$C_{\text{INST}@T=2.5}$} & \multicolumn{2}{c}{$C_{\text{AP}}$}  &  \multicolumn{2}{c}{$C_{\text{ERR}}$}\\
    \midrule
      $ A_{\text{ERG}}$* \FiveFlowerOpen & 0.643 & $ A_{\text{PE}}$ \SixFlowerPetalDotted & 0.651  &  $ A_{\text{PE}}$ \SixFlowerPetalDotted & 0.702 &   $ A_{\text{ETG}}$ \EightFlowerPetal & 0.688 & $ A_{\text{ERG}}$ * \EightFlowerPetal & 0.691 & $ A_{\text{ETG}}$  \EightFlowerPetal  & 0.587 \\
      
      $ A_{\text{ETG}}$  \FiveFlowerOpen & 0.643 & $ A_{\text{ERG}}$* \FiveFlowerOpen  & 0.644  & $ A_{\text{ERG}}$* \FiveFlowerOpen & 0.674 &  $ A_{\text{PE}}$ \SixFlowerAlternate & 0.673 &  $ A_{\text{ETG}}$ \SixFlowerAlternate & 0.655 & $ A_{\text{ERG}}$ \SixFlowerAlternate & 0.573\\
      
      $ A_{\text{avg}}$ \FiveFlowerOpen & 0.643 & $ A_{\text{ETG}}$ \FiveFlowerOpen  & 0.644 &  $ A_{\text{ETG}}$ \FiveFlowerOpen & 0.674 &   
      $ A_{\text{fin}}$ \FiveFlowerOpen  & 0.670 &  $ A_{\text{avg}}$ \SixFlowerAlternate & 0.641 & $ A_{\text{ERR}}$* \SixFlowerAlternate & 0.572\\
      
     $ A_{\text{max}}$ \FourClowerOpen & 0.490 & $ A_{\text{avg}}$ \FourClowerOpen & 0.638 & $ A_{\text{fin}}$ \FourClowerOpen & 0.667 &  
     $ A_{\text{ERG}}$* \FiveFlowerOpen & 0.669 & $ A_{\text{fin}}$ \OrnamentDiamondSolid & 0.568 & $ A_{\text{avg}}$ \FiveFlowerOpen & 0.544\\
     
     $ A_{\text{PE}}$  \OrnamentDiamondSolid & 0.430 & $ A_{\text{max}}$ & 0.614  &  $ A_{\text{avg}}$ \OrnamentDiamondSolid & 0.657 & 
     $ A_{\text{avg}}$ \FourClowerOpen  & 0.628 & $ A_{\text{PE}}$ \OrnamentDiamondSolid & 0.557 & $ A_{\text{fin}}$ \FourClowerOpen & 0.499\\
     
     $ A_{\text{fin}}$ & 0.276 & $ A_{\text{fin}}$ & 0.612  &  $ A_{\text{max}}$ & 0.639 & 
     $ A_{\text{max}}$ \OrnamentDiamondSolid & 0.619 &  $ A_{\text{max}}$ \OrnamentDiamondSolid & 0.546 & $ A_{\text{PE}}$ \OrnamentDiamondSolid & 0.478\\
     
     &   &   &     &   &   & $ A_{\text{ERR}}$ & 0.585 & $ A_{\text{ERR}}$ & 0.067 & $A_{\text{max}}$ & 0.448\\
  \bottomrule
\end{tabular}
\end{table*}
\begin{algorithm}
    \caption{Pseudocode for sampling a consistency $\tau$ score $B$ times for an evaluation metric $M$, given a set of runs for Topic Set $T$~\cite{Sakai-2021-ecir}. The function \texttt{truncate} returns the integer part of an argument.}
    \label{alg:const}
    \begin{algorithmic}[1]
        \State $n_1$ = \texttt{truncate}($\left| T \right|/2$), $n_2$ = $\left| T \right| - n_1$
        \ForTo{$b = 1$}{$B$}
            \State $T^b_1$ = a random subset of the original topic set $T$ \text{ s.t.} $\left|T_1 \right|$ = $n_1$
            \State $T^b_2 = T - T_1$ \Comment{$\left| T_2 \right| = n_2$}
            \State $r^b_1$ = a list of run rankings according to mean $M$ over $T^b_1$
            \State  $r^b_2$= a list of run rankings according to mean $M$ over $T^b_2$
            \State $\tau^{b}$ = Kendall’s $\tau$ score for run rankings $r^b_1$ and $r^b_2$
        \EndFor
    \end{algorithmic}
\end{algorithm}

This section compares the performance of different aggregations given the browsing model of a metric in terms of system ranking consistency across two disjoint topic sets. To be more specific, given a test collection whose topic set is $T$ and a set of $K$ runs associated with it, we compare a set $\{M\}$ of candidate metrics following the method in previous work ~\cite{Sakai-2021-ecir} as follows. 

\begin{enumerate}
\item For each measure $M$, evaluate the $K$ runs with $T$, and thereby obtain a $\left|T\right| \times K$ topic-by-run score matrix $S_M$ .
\item From each  $S_M$, obtain a $\tau$ score $B$ times using the algorithm shown in Algorithm ~\ref{alg:const}, where each $\tau$ quantifies the system ranking consistency when the K runs are ranked according to two disjoint subsets of $T$ . We thus obtain a $B \times \left|\{M\}\right|$ matrix $C$ containing the consistency $\tau$ scores.
\item To see if any of the differences in mean consistency $\tau$ scores are statistically significant, apply a paired, randomised Tukey HSD test~\cite{Carterette-2012, Sakai-2018} to $C$.
\end{enumerate}

 Note that Tukey HSD test is a multiple comparison procedure, one can thus ensure that the familywise Type I error rate is no more than $\alpha$, which is set to $0.05$ throughout our study. Moreover, as the randomised Tukey HSD test is distribution-free, it can be applied regardless the distribution of $\tau$ scores. We use the \texttt{Random-test} script of the \texttt{Discpower} tool~\footnote{http://research.nii.ac.jp/ntcir/tools/discpower-en.html} for the randomised Tukey HSD test with $2,000$ trials.

Table ~\ref{tab:const} summarises the results of our system ranking consistency experiments with $B = 1, 000$ topic subset pairs in each case. From Table ~\ref{tab:const} we can observe the following result.

In general, $A_{\text{ERG}}$ and $A_{\text{ETG}}$ performs well in terms of system ranking consistency. $A_{\text{ERG}}$ has the best or the second best system ranking consistency in all but one cases ~(INST on TR19DL). $A_{\text{ETG}}$ also has the best or the second best system ranking consistency in all but two cases with two exceptions~(INST and ERR on WWW3). 

$A_{\text{max}}$ tends to return system rankings with discrepancy on different topic sets. It is in the last place or the penultimate in terms of system ranking consistency in all but one cases~(Precision on TR19DL). A possible explanation for the low consistency of system rankings given by $A_{\text{max}}$ is that, it always returns the same result after it encounters the maximum relevance score so far, and thus the metric scores tend to be similar, which impairs the ability of the metric score to discriminate runs. 

The performance of $A_{\text{avg}}$ in terms of system ranking consistency is mediocre while stable. Overall, the trend is that it underperforms $A_{\text{ERG}}$ and $A_{\text{ETG}}$ while outperforming $A_{\text{max}}$ in terms of system ranking consistency. 

The performance of $A_{\text{fin}}$ in terms of system ranking consistency is volatile. In general, two trends can be observed:~(1)~It does not perform well in terms of system ranking consistency under the browsing model of Precision, where it is in the last place.~(2)~Its performance in terms of system ranking consistency is mediocre under the browsing model of ERR, where it is in the third last place. However, it is hard to summarise its performance briefly when it is combined with the browsing model of other metrics, as its performance varies among different datasets. For example, on WWW3, it has the third-best system ranking consistency under the browsing model of DCG, but on TR19DL, likewise under the browsing model of DCG, it is in the last place. 

The performance of $A_{\text{PE}}$ in terms of system ranking consistency is also unstable. Under the browsing model of AP and ERR, its performance tends to be the
the compromise of $A_{\text{fin}}$ and $A_{\text{max}}$. However, in other cases, this trend cannot be confirmed. What we observed is that, on one dataset its ranking in system ranking consistency is between $A_{\text{fin}}$ and $A_{\text{max}}$, but on the other it ranks higher than both.

$A_{\text{ERR}}$ has a outstanding performance in system ranking consistency under the browsing model of ERR, but its performance falters when it is combined with the browsing model of INST and AP. This result suggests that $A_{\text{ERR}}$ is a highly specialised aggregation function for the browsing model of ERR and might have a bad performance in terms of system ranking consistency when it is used for the model of other metrics.

From the perspective of canonical and alternative aggregations, the overall picture is that metrics with their canonical aggregation all have favourable performances in system ranking consistency. For ERR, replacing $A_{\text{ERR}}$ with $A_{\text{ERG}}$ might further drive up its performance in terms of system ranking consistency, but whether the improvement is substantial needs further verification in future work. Current result shows that the improvement on WWW3 dataset is statistically significant while the improvement on TR19DL dataset is incremental and lacks statistical significance.
\section{Discriminative Power}
\label{sec:disc}

In offline evaluation practice, a metric that tends to significantly discriminate more system pairs is preferred. The ability to significantly discriminate system pairs is called \textit{discriminative power}. To figure out the discriminative power of the metrics, we compute the scores of each metric for $K$ runs on $\left|T\right|$ topics with cutoff $L$ = 10. Thus, for each metric, we have a $\left|T\right| \times K$ score matrix and we have $K * (K - 1) / 2$ system pairs on $\left|T\right|$ topics. We then carry out significance tests for the difference of metric scores on each system pair. For significance testing, we used the randomised version of the paired Tukey HSD test, using the Discpower tool with $2,000$ trials. The algorithm to obtain Achieved Significance Level (ASL) is given by the algorithm in the work of Carterette~\cite{Carterette-2012}.

Figure~\ref{fig:asl} shows the result in the form of ASL curve. Metrics whose curves are close to the origin are the ones with high discriminative power, which means that they produce smaller p-values for many run pairs than other metrics do. Note that for Precision, DCG and RBP, the ASL curves of $A_{\text{ETG}}$ are not shown in the figures as they are exactly the same as the ASL curves of $A_{\text{ERG}}$, since the metric score given by $A_{\text{ETG}}$ is equal to the metric score given by $A_{\text{ERG}}$ multiplying with a constant. For  Precision, the ASL curves of $A_{\text{avg}}$ are not shown in the figure as it is exactly the same as the ASL curves of $A_{\text{ERG}}$ and $A_{\text{ETG}}$, since the metric score given by $A_{\text{avg}}$ is equal to the metric score given by $A_{\text{ETG}}$ multiplying with a constant. From Figure~\ref{fig:asl}, we can observe the following result.

$A_{\text{ERG}}$ brings a strong discriminative power and in general performs superbly among different metrics. $A_{\text{ETG}}$ also has a hefty discriminative power in most cases, but it hobbles when it is combined with the browsing model of ERR. Under the browsing model of  Precision, DCG and RBP, it has same ASL curves as the one of $A_{\text{ERG}}$ and it is thus the one of best performers in discriminative power. Under the browsing model of INST and AP, the discriminative power of $A_{\text{ETG}}$ is also strong, akin to $A_{\text{ERG}}$. Nevertheless, the discriminative power of $A_{\text{ETG}}$ tones down under the browsing model of ERR, being dwarfed by $A_{\text{avg}}$ and the canonical $A_{\text{ERR}}$. Considering the fact that $A_{\text{ETG}}(i)$ is equal to $V^+ \cdot A_{\text{ERG}}(i)$, the dismal performance of $A_{\text{ETG}}$ might have potential relation with the volatile browsing model of ERR. 

$A_{\text{max}}$ has a frail discriminative power and is a lagger in general. It is in the last place in terms of discriminative power under the browsing model of DCG, RBP, INST and ERR. It is in the penultimate place under the browsing model of AP, only outstripping $A_{\text{ERR}}$, whose discriminative power is tenuous in that case. Similar to what causes the low consistency of system rankings given by $A_{\text{max}}$ the possible explanation for the weak discriminative power of $A_{\text{max}}$ is that, it always returns the same result after it encounters the maximum relevance score so far, and thus the metric scores tend to be similar, which impairs the ability of the metric score to discriminate runs. 

The discriminative power of $A_{\text{avg}}$ is mediocre while stable, just like its performance in terms of system ranking consistency. In general, it is dwarfed by $A_{\text{ERG}}$ and $A_{\text{ERG}}$ while outstripping $A_{\text{max}}$ in discriminative power.  

The discriminative power of $A_{\text{fin}}$ is volatile and highly depend on the browsing model of a metric. Under the browsing model of RBP and INST, it has a strong discriminative power, performing similar to or even better than $A_{\text{ERG}}$ and $A_{\text{ETG}}$. Under the browsing model of AP and DCG, its discriminative power is mediocre. Its discriminative power falters under the browsing model of Precision, where it is in the last place in terms of discriminative power. 

The discriminative power of $A_{\text{PE}}$ is prone to be the compromise of $A_{\text{fin}}$ and $A_{\text{max}}$ in general. This result is intuitive if one considers the definition of  $A_{\text{PE}}$. The only exception is that, on WWW3 dataset it outperforms both $A_{\text{fin}}$ and $A_{\text{max}}$ in discriminative power when it is combined with the browsing model of Precision.

$A_{\text{ERR}}$ performs well in discriminative power under the browsing model of ERR, merely being inferior to $A_{\text{ERG}}$. Nevertheless, it has an insufficient performance under the browsing model of AP and INST. Especially in the case of AP, its discriminative power is substantially weaker than other aggregations. This result again suggests that $A_{\text{ERR}}$ is a highly specialised aggregation function for the browsing model of ERR and might perform poorly in terms of discriminative power when being combined with the browsing model of other metrics.

From the perspective of canonical and alternative aggregations, the overall picture is that metrics with their canonical aggregation all have good, if not the best, performances in discriminative power. Nevertheless, for ERR, replacing $A_{\text{ERR}}$ with $A_{\text{ERG}}$ can further strengthen the discriminative power. 

\begin{figure*}[htbp]
\centering
\begin{subfigure}[t]{0.33\textwidth}
    \centering
    \includegraphics[width=5.2cm]{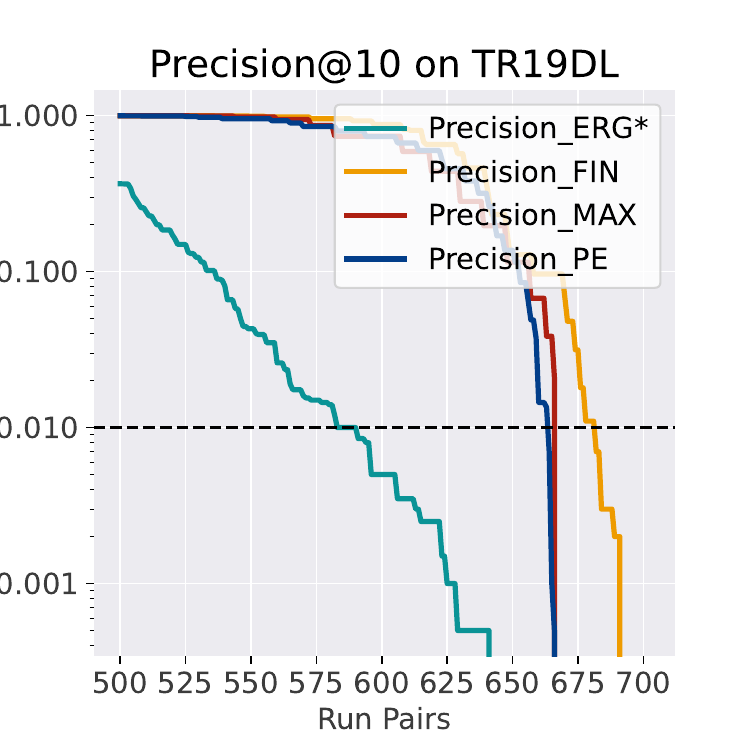}
    \caption*{}
\end{subfigure} 
\begin{subfigure}[t]{0.33\textwidth}
    \centering
    \includegraphics[width=5.2cm]{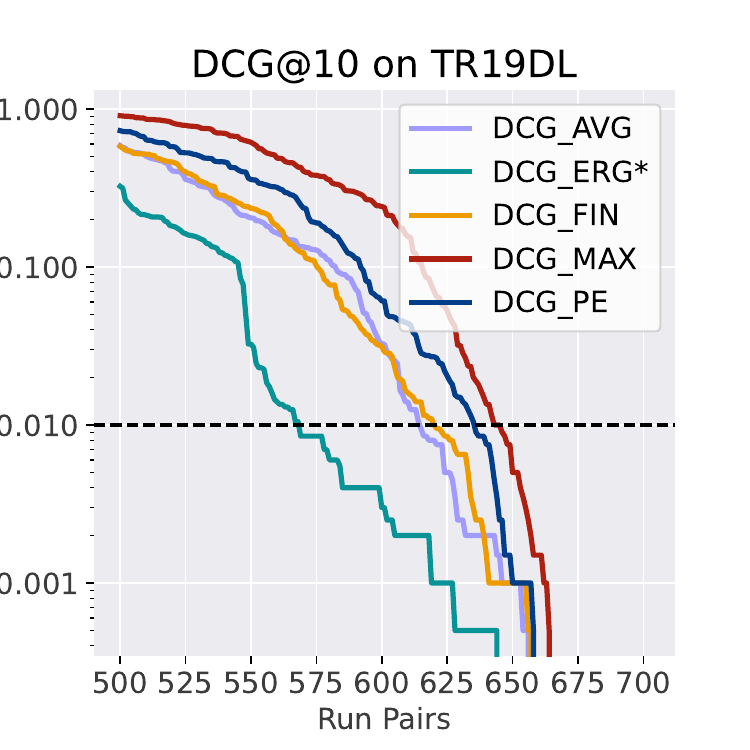}
    \caption*{}
\end{subfigure}
\begin{subfigure}[t]{0.33\textwidth}
    \centering
    \includegraphics[width=5.2cm]{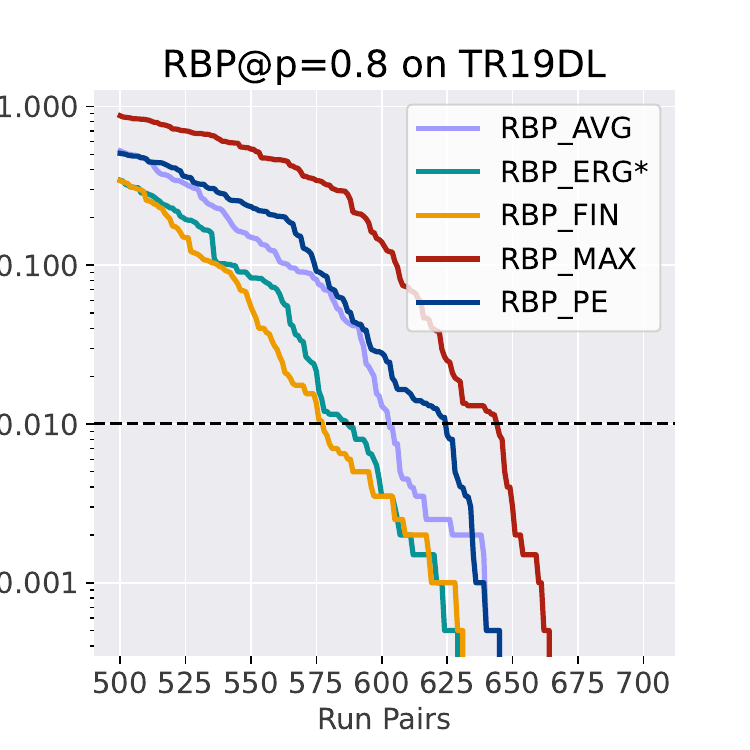}
    \caption*{}
\end{subfigure}
\begin{subfigure}[t]{0.33\textwidth}
    \centering
    \includegraphics[width=5.2cm]{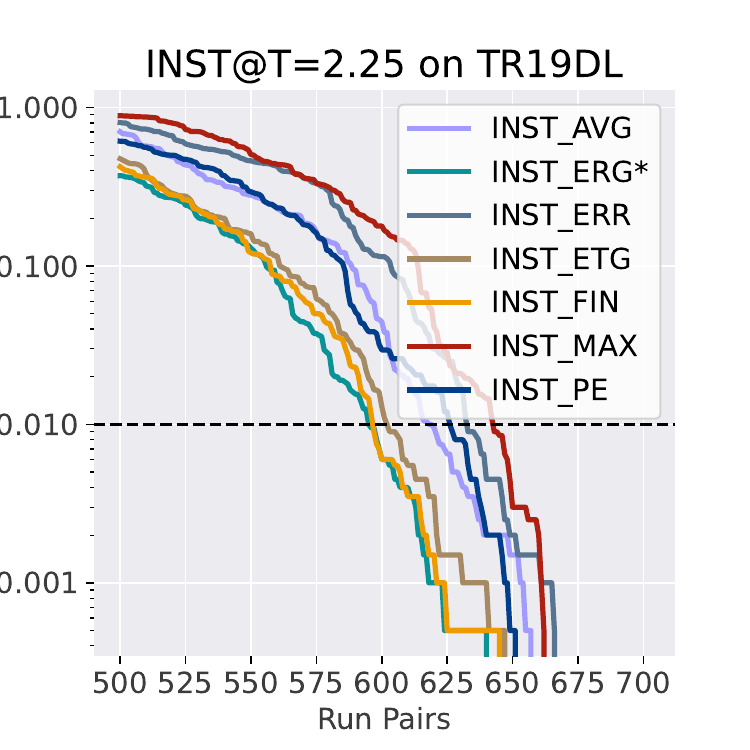}
    \caption*{}
\end{subfigure}
\begin{subfigure}[t]{0.33\textwidth}
    \centering
    \includegraphics[width=5.2cm]{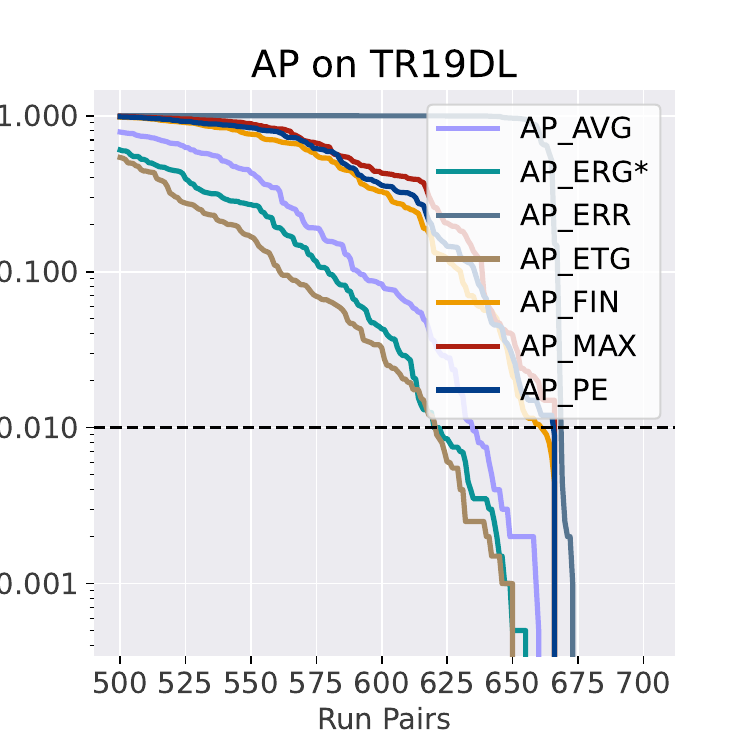}
    \caption*{}
\end{subfigure}
\begin{subfigure}[t]{0.33\textwidth}
    \centering
    \includegraphics[width=5.2cm]{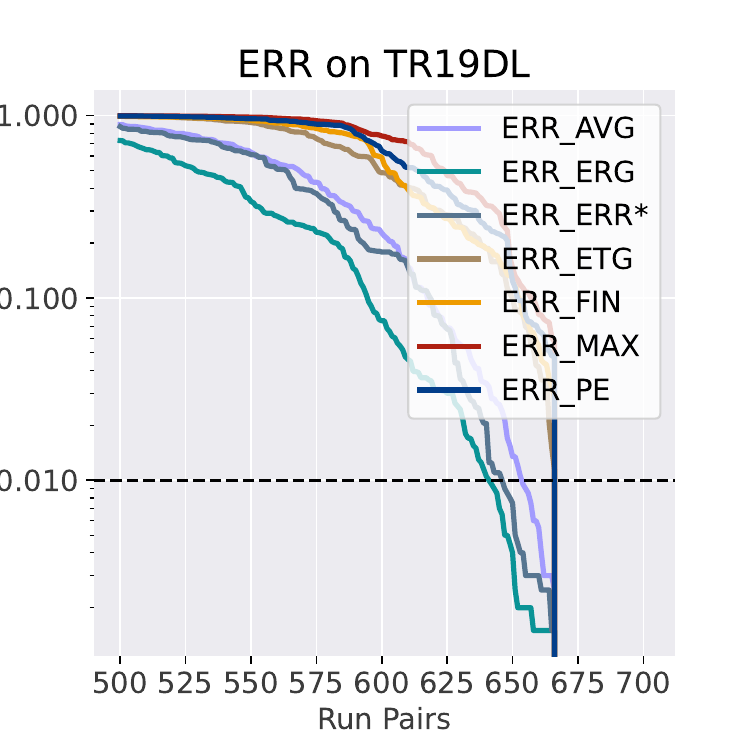}
    \caption*{}
\end{subfigure}
\begin{subfigure}[t]{0.33\textwidth}
    \centering
    \includegraphics[width=5.2cm]{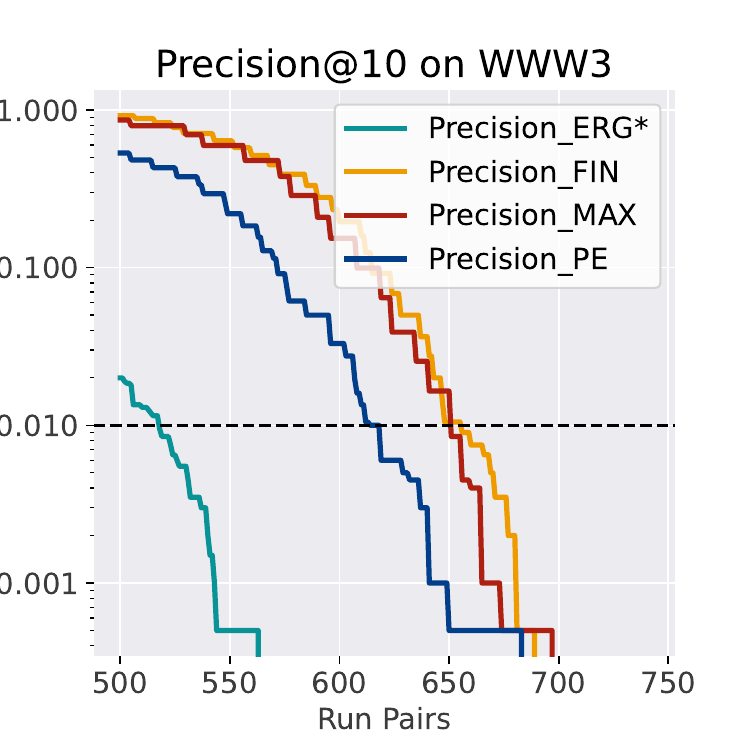}
    \caption*{}
\end{subfigure}
\begin{subfigure}[t]{0.33\textwidth}
    \centering
    \includegraphics[width=5.2cm]{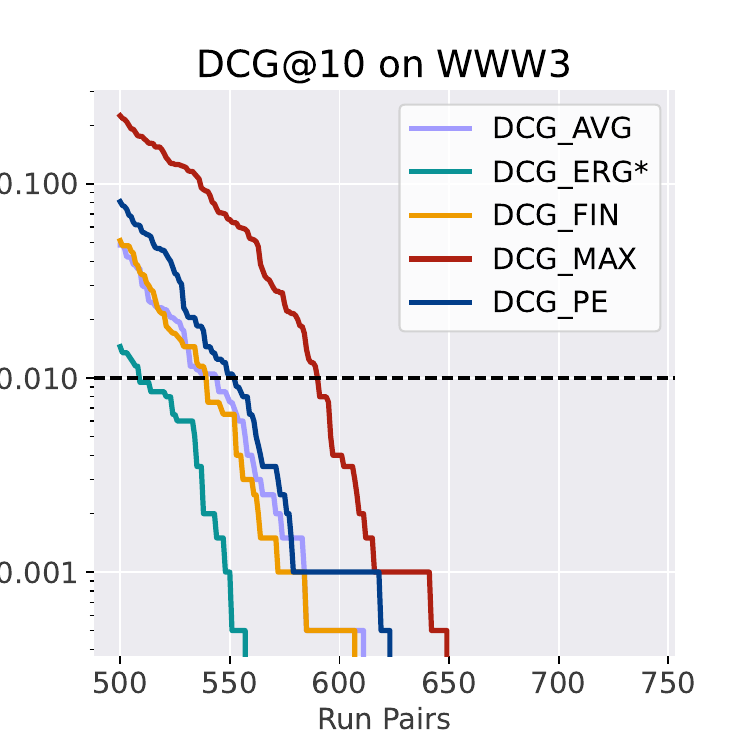}
    \caption*{}
\end{subfigure}
\begin{subfigure}[t]{0.33\textwidth}
    \centering
    \includegraphics[width=5.2cm]{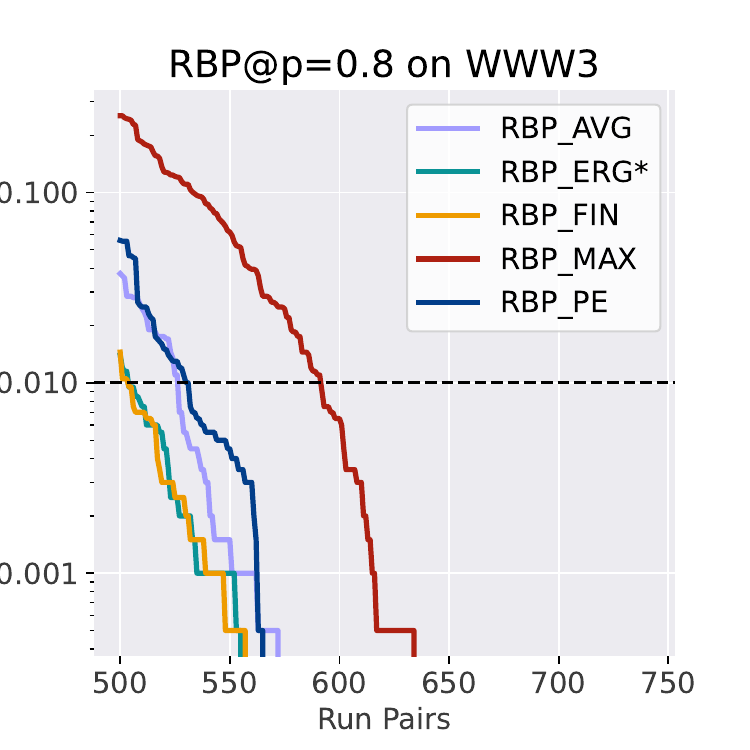}
    \caption*{}
\end{subfigure}
\end{figure*}
\clearpage

\begin{figure*}[htbp]\ContinuedFloat
\centering
\begin{subfigure}[t]{0.33\textwidth}
    \centering
    \includegraphics[width=5.2cm]{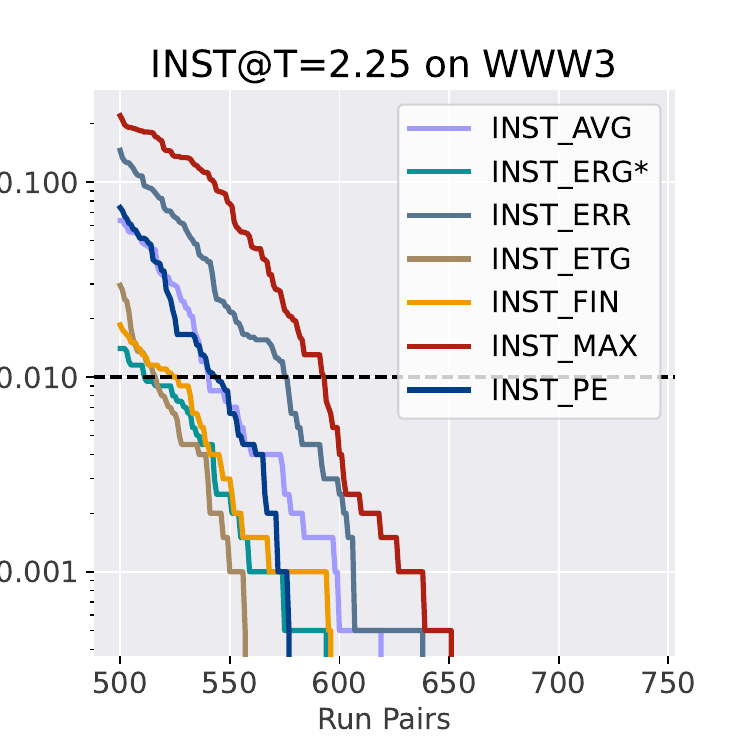}
    \caption*{}
\end{subfigure}
\begin{subfigure}[t]{0.33\textwidth}
    \centering
    \includegraphics[width=5.2cm]{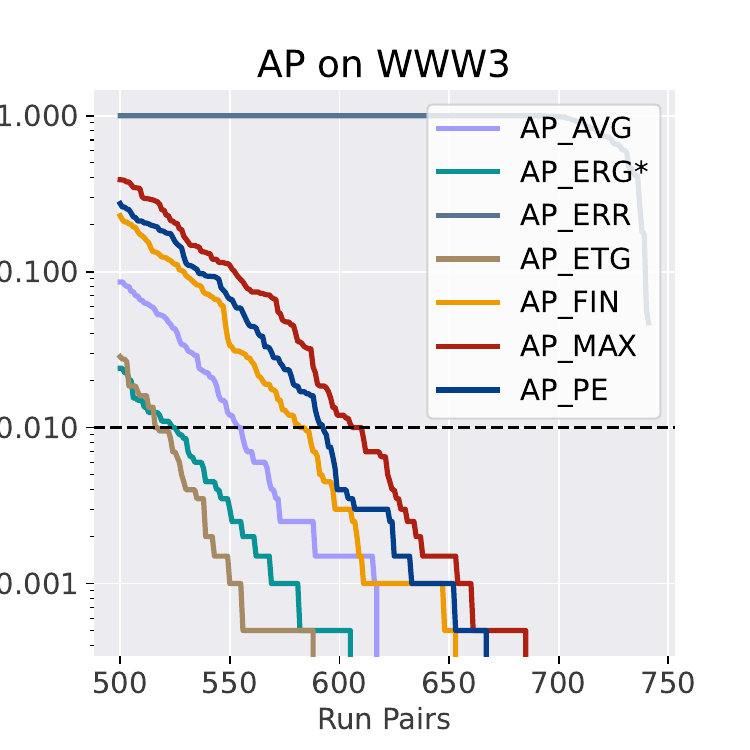}
    \caption*{}
\end{subfigure}
\begin{subfigure}[t]{0.33\textwidth}
    \centering
    \includegraphics[width=5.2cm]{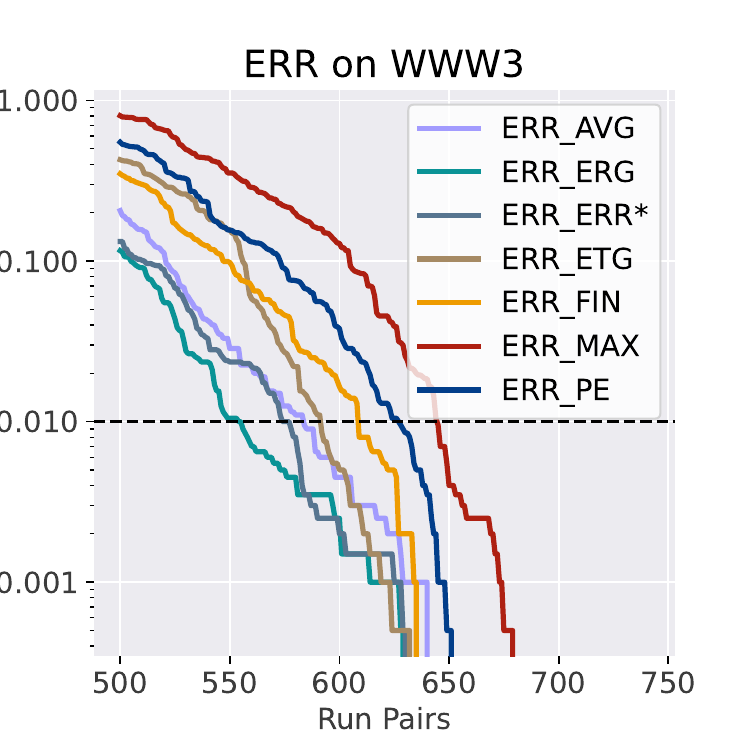}
    \caption*{}
\end{subfigure}

\caption{ASL Curves of C/W/L/A Metrics on WWW3 collection and TR19DL collection respectively. The asterisk indicates the canonical aggregation of each metric.}
\Description{ASL Curves of C/W/L/A Metrics on WWW3 collection and TR19DL collection respectively. The asterisk indicates the canonical aggregation of each metric.}
\label{fig:asl}
\end{figure*}

\section{Conclusions and Discussion}
In this study, we meta-evaluated metrics obtained by combining different aggregation functions with the browsing model of Precision, DCG, RBP, INST, AP and ERR. We compared these metrics in order to figure out that:~given the browsing model of a metric, what is the impact of using different aggregation functions on system ranking similarity, system ranking consistency and discriminative power. Our work extends the work of Moffat \textit{et al.}~\cite{Moffat-2022} from the perspective of statistical reliability in offline evaluation experiment. Our experimental results provide a useful insight for researchers who are going to design reliable evaluation metrics for offline evaluation using the C/W/L/A framework. With respect to the RQs, we have the following findings:

\textbf{RQ1:~The system ranking similarity among aggregations}. The system ranking similarity among different aggregations depends on the browsing model of a metric and it is hard to give a universal rule. 

\textbf{RQ2:~The system ranking consistency of aggregations}. $A_{\text{ERG}}$ and $A_{\text{ETG}}$ have outstanding performances in terms of system ranking consistency. $A_{\text{max}}$ usually performs poorly in terms of system ranking consistency. The performance of $A_{\text{avg}}$ in terms of system ranking consistency is mediocre. The performances of $A_{\text{fin}}$ and $A_{\text{PE}}$ in terms of system ranking consistency are volatile, depending on the browsing model of a metric. $A_{\text{ERR}}$ has a outstanding performance in system ranking consistency under the browsing model of ERR, but it performs poorly when being combined with the browsing model of INST and AP.

\textbf{RQ3:~The discriminative power of aggregations}. $A_{\text{ERG}}$ tends to have the strongest discriminative power and performs the best in most cases. $A_{\text{ETG}}$ also has outstanding performance in terms of discriminative power except for the case of ERR. $A_{\text{max}}$ tends to have a weak discriminative power and has an insufficient performance in most cases. The discriminative power of $A_{\text{avg}}$ is mediocre. The discriminative power of $A_{\text{fin}}$ is volatile and highly depend on the browsing model of a metric. The discriminative power of $A_{\text{PE}}$ is prone to be the compromise of $A_{\text{fin}}$ and $A_{\text{max}}$ in most cases. $A_{\text{ERR}}$ performs well in discriminative power under the browsing model of ERR, but it performs poorly under the browsing model of AP and INST. 

\textbf{RQ4:~Alternative aggregations that improve the statistical reliability of metrics}. Given that the canonical aggregation of Precision, DCG, RBP, INST and AP is $A_{\text{ERG}}$, and that $A_{\text{ERG}}$ has been mentioned above as performing well in terms of system ranking consistency and discriminative power, there is no evidence that replacing the canonical aggregation with alternative aggregation would further improve their performance. For ERR, replacing $A_{\text{ERR}}$ with the canonical $A_{\text{ERG}}$ can further strengthen the discriminative power while obtaining a system ranking list similar to the canonical version.

Overall, our result suggests that, in terms of system ranking consistency and discriminative power, $A_{\text{ERG}}$ has an outstanding performance while $A_{\text{max}}$ usually has an insufficient performance. A possible explanation is that:~$A_{\text{ERG}}$ uses the information of all relevance scores it has encountered so far, while using the information of the probability of users inspecting documents on each rank~($1/V^+$). Therefore, metric scores given by $A_{\text{ERG}}$ are able to discriminate more runs. On the other hand, $A_{\text{max}}$ only uses the information of the maximum relevance score it has encountered so far, and thus the metric scores tend to be similar, which impairs the ability of the metric score to discriminate runs. 

Based on the results in this study, we recommend IR researchers to:~(1)~use ERR with $A_{\text{ERG}}$ in offline evaluation practice in order to achieve high system ranking consistency and discriminative power while obtaining a system ranking list similar to the canonical version at the same time;~(2)~use $A_{\text{ERG}}$ as the aggregation function when designing evaluation metrics using the C/W/L/A framework. This is conducive to improve the system ranking consistency and discriminative power of the metrics.


\bibliographystyle{ACM-Reference-Format}
\balance
\bibliography{main}


\begin{thebibliography}{40}


\ifx \showCODEN    \undefined \def \showCODEN     #1{\unskip}     \fi
\ifx \showDOI      \undefined \def \showDOI       #1{#1}\fi
\ifx \showISBNx    \undefined \def \showISBNx     #1{\unskip}     \fi
\ifx \showISBNxiii \undefined \def \showISBNxiii  #1{\unskip}     \fi
\ifx \showISSN     \undefined \def \showISSN      #1{\unskip}     \fi
\ifx \showLCCN     \undefined \def \showLCCN      #1{\unskip}     \fi
\ifx \shownote     \undefined \def \shownote      #1{#1}          \fi
\ifx \showarticletitle \undefined \def \showarticletitle #1{#1}   \fi
\ifx \showURL      \undefined \def \showURL       {\relax}        \fi
\providecommand\bibfield[2]{#2}
\providecommand\bibinfo[2]{#2}
\providecommand\natexlab[1]{#1}
\providecommand\showeprint[2][]{arXiv:#2}

\bibitem[\protect\citeauthoryear{Al-Maskari, Sanderson, Clough, and
  Airio}{Al-Maskari et~al\mbox{.}}{2008}]%
        {maskari-2008}
\bibfield{author}{\bibinfo{person}{Azzah Al-Maskari}, \bibinfo{person}{Mark
  Sanderson}, \bibinfo{person}{Paul Clough}, {and} \bibinfo{person}{Eija
  Airio}.} \bibinfo{year}{2008}\natexlab{}.
\newblock \showarticletitle{The Good and the Bad System: Does the Test
  Collection Predict Users' Effectiveness?}. In
  \bibinfo{booktitle}{\emph{Proceedings of the 31st Annual International ACM
  SIGIR Conference on Research and Development in Information Retrieval}}
  (Singapore, Singapore) \emph{(\bibinfo{series}{SIGIR '08})}.
  \bibinfo{publisher}{Association for Computing Machinery},
  \bibinfo{address}{New York, NY, USA}, \bibinfo{pages}{59–66}.
\newblock
\showISBNx{9781605581644}
\urldef\tempurl%
\url{https://doi.org/10.1145/1390334.1390347}
\showDOI{\tempurl}


\bibitem[\protect\citeauthoryear{Amigo, Gonzalo, Mizzaro, and Carrillo-de
  Albornoz}{Amigo et~al\mbox{.}}{2020}]%
        {amigo-etal-2020-effectiveness}
\bibfield{author}{\bibinfo{person}{Enrique Amigo}, \bibinfo{person}{Julio
  Gonzalo}, \bibinfo{person}{Stefano Mizzaro}, {and} \bibinfo{person}{Jorge
  Carrillo-de Albornoz}.} \bibinfo{year}{2020}\natexlab{}.
\newblock \showarticletitle{An Effectiveness Metric for Ordinal Classification:
  Formal Properties and Experimental Results}. In
  \bibinfo{booktitle}{\emph{Proceedings of the 58th Annual Meeting of the
  Association for Computational Linguistics}}. \bibinfo{publisher}{Association
  for Computational Linguistics}, \bibinfo{address}{Online},
  \bibinfo{pages}{3938--3949}.
\newblock
\urldef\tempurl%
\url{https://doi.org/10.18653/v1/2020.acl-main.363}
\showDOI{\tempurl}


\bibitem[\protect\citeauthoryear{Amig\'{o}, Spina, and Carrillo-de
  Albornoz}{Amig\'{o} et~al\mbox{.}}{2018}]%
        {Amigo-2018}
\bibfield{author}{\bibinfo{person}{Enrique Amig\'{o}}, \bibinfo{person}{Damiano
  Spina}, {and} \bibinfo{person}{Jorge Carrillo-de Albornoz}.}
  \bibinfo{year}{2018}\natexlab{}.
\newblock \showarticletitle{An Axiomatic Analysis of Diversity Evaluation
  Metrics: Introducing the Rank-Biased Utility Metric}. In
  \bibinfo{booktitle}{\emph{The 41st International ACM SIGIR Conference on
  Research \& Development in Information Retrieval}} (Ann Arbor, MI, USA)
  \emph{(\bibinfo{series}{SIGIR '18})}. \bibinfo{publisher}{Association for
  Computing Machinery}, \bibinfo{address}{New York, NY, USA},
  \bibinfo{pages}{625–634}.
\newblock
\showISBNx{9781450356572}
\urldef\tempurl%
\url{https://doi.org/10.1145/3209978.3210024}
\showDOI{\tempurl}


\bibitem[\protect\citeauthoryear{Anelli, Di~Noia, Di~Sciascio, Pomo, and
  Ragone}{Anelli et~al\mbox{.}}{2019}]%
        {Anelli-2019}
\bibfield{author}{\bibinfo{person}{Vito~Walter Anelli},
  \bibinfo{person}{Tommaso Di~Noia}, \bibinfo{person}{Eugenio Di~Sciascio},
  \bibinfo{person}{Claudio Pomo}, {and} \bibinfo{person}{Azzurra Ragone}.}
  \bibinfo{year}{2019}\natexlab{}.
\newblock \showarticletitle{On the Discriminative Power of Hyper-Parameters in
  Cross-Validation and How to Choose Them}. In
  \bibinfo{booktitle}{\emph{Proceedings of the 13th ACM Conference on
  Recommender Systems}} (Copenhagen, Denmark) \emph{(\bibinfo{series}{RecSys
  '19})}. \bibinfo{publisher}{Association for Computing Machinery},
  \bibinfo{address}{New York, NY, USA}, \bibinfo{pages}{447–451}.
\newblock
\showISBNx{9781450362436}
\urldef\tempurl%
\url{https://doi.org/10.1145/3298689.3347010}
\showDOI{\tempurl}


\bibitem[\protect\citeauthoryear{Azzopardi, Mackenzie, and Moffat}{Azzopardi
  et~al\mbox{.}}{2021}]%
        {Azzopardi-2021-ERR}
\bibfield{author}{\bibinfo{person}{Leif Azzopardi}, \bibinfo{person}{Joel
  Mackenzie}, {and} \bibinfo{person}{Alistair Moffat}.}
  \bibinfo{year}{2021}\natexlab{}.
\newblock \showarticletitle{ERR is Not C/W/L: Exploring the Relationship
  Between Expected Reciprocal Rank and Other Metrics}. In
  \bibinfo{booktitle}{\emph{Proceedings of the 2021 ACM SIGIR International
  Conference on Theory of Information Retrieval}} (Virtual Event, Canada)
  \emph{(\bibinfo{series}{ICTIR '21})}. \bibinfo{publisher}{Association for
  Computing Machinery}, \bibinfo{address}{New York, NY, USA},
  \bibinfo{pages}{231–237}.
\newblock
\showISBNx{9781450386111}
\urldef\tempurl%
\url{https://doi.org/10.1145/3471158.3472239}
\showDOI{\tempurl}


\bibitem[\protect\citeauthoryear{Azzopardi, Thomas, and Craswell}{Azzopardi
  et~al\mbox{.}}{2018}]%
        {Azzopardi-2018-IFT}
\bibfield{author}{\bibinfo{person}{Leif Azzopardi}, \bibinfo{person}{Paul
  Thomas}, {and} \bibinfo{person}{Nick Craswell}.}
  \bibinfo{year}{2018}\natexlab{}.
\newblock \showarticletitle{Measuring the Utility of Search Engine Result
  Pages: An Information Foraging Based Measure}. In
  \bibinfo{booktitle}{\emph{The 41st International ACM SIGIR Conference on
  Research {\&} Development in Information Retrieval}} (Ann Arbor, MI, USA)
  \emph{(\bibinfo{series}{SIGIR '18})}. \bibinfo{publisher}{Association for
  Computing Machinery}, \bibinfo{address}{New York, NY, USA},
  \bibinfo{pages}{605–614}.
\newblock
\showISBNx{9781450356572}
\urldef\tempurl%
\url{https://doi.org/10.1145/3209978.3210027}
\showDOI{\tempurl}


\bibitem[\protect\citeauthoryear{Azzopardi, White, Thomas, and
  Craswell}{Azzopardi et~al\mbox{.}}{2020}]%
        {azzopardi-2020-data}
\bibfield{author}{\bibinfo{person}{Leif Azzopardi}, \bibinfo{person}{Ryen~W.
  White}, \bibinfo{person}{Paul Thomas}, {and} \bibinfo{person}{Nick
  Craswell}.} \bibinfo{year}{2020}\natexlab{}.
\newblock \showarticletitle{Data-Driven Evaluation Metrics for Heterogeneous
  Search Engine Result Pages} \emph{(\bibinfo{series}{CHIIR '20})}.
  \bibinfo{publisher}{Association for Computing Machinery},
  \bibinfo{address}{New York, NY, USA}, \bibinfo{pages}{213–222}.
\newblock
\showISBNx{9781450368926}
\urldef\tempurl%
\url{https://doi.org/10.1145/3343413.3377959}
\showDOI{\tempurl}


\bibitem[\protect\citeauthoryear{Bailey, Moffat, Scholer, and Thomas}{Bailey
  et~al\mbox{.}}{2015}]%
        {Bailey-2015}
\bibfield{author}{\bibinfo{person}{Peter Bailey}, \bibinfo{person}{Alistair
  Moffat}, \bibinfo{person}{Falk Scholer}, {and} \bibinfo{person}{Paul
  Thomas}.} \bibinfo{year}{2015}\natexlab{}.
\newblock \showarticletitle{User Variability and IR System Evaluation}. In
  \bibinfo{booktitle}{\emph{Proceedings of the 38th International ACM SIGIR
  Conference on Research and Development in Information Retrieval}} (Santiago,
  Chile) \emph{(\bibinfo{series}{SIGIR '15})}. \bibinfo{publisher}{Association
  for Computing Machinery}, \bibinfo{address}{New York, NY, USA},
  \bibinfo{pages}{625–634}.
\newblock
\showISBNx{9781450336215}
\urldef\tempurl%
\url{https://doi.org/10.1145/2766462.2767728}
\showDOI{\tempurl}


\bibitem[\protect\citeauthoryear{Buckley and Voorhees}{Buckley and
  Voorhees}{2000}]%
        {Buckley-2000}
\bibfield{author}{\bibinfo{person}{Chris Buckley} {and}
  \bibinfo{person}{Ellen~M. Voorhees}.} \bibinfo{year}{2000}\natexlab{}.
\newblock \showarticletitle{Evaluating Evaluation Measure Stability}. In
  \bibinfo{booktitle}{\emph{Proceedings of the 23rd Annual International ACM
  SIGIR Conference on Research and Development in Information Retrieval}}
  (Athens, Greece) \emph{(\bibinfo{series}{SIGIR '00})}.
  \bibinfo{publisher}{Association for Computing Machinery},
  \bibinfo{address}{New York, NY, USA}, \bibinfo{pages}{33–40}.
\newblock
\showISBNx{1581132263}
\urldef\tempurl%
\url{https://doi.org/10.1145/345508.345543}
\showDOI{\tempurl}


\bibitem[\protect\citeauthoryear{B{\"u}ttcher, Clarke, Yeung, and
  Soboroff}{B{\"u}ttcher et~al\mbox{.}}{2007}]%
        {Bttcher-2007}
\bibfield{author}{\bibinfo{person}{Stefan B{\"u}ttcher},
  \bibinfo{person}{Charles L.~A. Clarke}, \bibinfo{person}{Peter C.~K. Yeung},
  {and} \bibinfo{person}{Ian Soboroff}.} \bibinfo{year}{2007}\natexlab{}.
\newblock \showarticletitle{Reliable information retrieval evaluation with
  incomplete and biased judgements}. In \bibinfo{booktitle}{\emph{SIGIR}}.
\newblock


\bibitem[\protect\citeauthoryear{Carterette}{Carterette}{2012}]%
        {Carterette-2012}
\bibfield{author}{\bibinfo{person}{Benjamin~A. Carterette}.}
  \bibinfo{year}{2012}\natexlab{}.
\newblock \showarticletitle{Multiple Testing in Statistical Analysis of
  Systems-Based Information Retrieval Experiments}.
\newblock \bibinfo{journal}{\emph{ACM Trans. Inf. Syst.}} \bibinfo{volume}{30},
  \bibinfo{number}{1}, Article \bibinfo{articleno}{4} (\bibinfo{date}{mar}
  \bibinfo{year}{2012}), \bibinfo{numpages}{34}~pages.
\newblock
\showISSN{1046-8188}
\urldef\tempurl%
\url{https://doi.org/10.1145/2094072.2094076}
\showDOI{\tempurl}


\bibitem[\protect\citeauthoryear{Chapelle, Metlzer, Zhang, and
  Grinspan}{Chapelle et~al\mbox{.}}{2009}]%
        {Chapelle-2009}
\bibfield{author}{\bibinfo{person}{Olivier Chapelle}, \bibinfo{person}{Donald
  Metlzer}, \bibinfo{person}{Ya Zhang}, {and} \bibinfo{person}{Pierre
  Grinspan}.} \bibinfo{year}{2009}\natexlab{}.
\newblock \showarticletitle{Expected Reciprocal Rank for Graded Relevance}. In
  \bibinfo{booktitle}{\emph{Proceedings of the 18th ACM Conference on
  Information and Knowledge Management}} (Hong Kong, China)
  \emph{(\bibinfo{series}{CIKM '09})}. \bibinfo{publisher}{Association for
  Computing Machinery}, \bibinfo{address}{New York, NY, USA},
  \bibinfo{pages}{621–630}.
\newblock
\showISBNx{9781605585123}
\urldef\tempurl%
\url{https://doi.org/10.1145/1645953.1646033}
\showDOI{\tempurl}


\bibitem[\protect\citeauthoryear{Chen, Zhou, Liu, Zhang, and Ma}{Chen
  et~al\mbox{.}}{2017}]%
        {Chen-2017-THUIR1}
\bibfield{author}{\bibinfo{person}{Ye Chen}, \bibinfo{person}{Ke Zhou},
  \bibinfo{person}{Yiqun Liu}, \bibinfo{person}{Min Zhang}, {and}
  \bibinfo{person}{Shaoping Ma}.} \bibinfo{year}{2017}\natexlab{}.
\newblock \showarticletitle{Meta-Evaluation of Online and Offline Web Search
  Evaluation Metrics} \emph{(\bibinfo{series}{SIGIR '17})}.
  \bibinfo{publisher}{Association for Computing Machinery},
  \bibinfo{address}{New York, NY, USA}, \bibinfo{pages}{15–24}.
\newblock
\showISBNx{9781450350228}
\urldef\tempurl%
\url{https://doi.org/10.1145/3077136.3080804}
\showDOI{\tempurl}


\bibitem[\protect\citeauthoryear{Fredrickson and Kahneman}{Fredrickson and
  Kahneman}{1993}]%
        {Fredrickson-1993-Duration}
\bibfield{author}{\bibinfo{person}{Barbara~L. Fredrickson} {and}
  \bibinfo{person}{Daniel Kahneman}.} \bibinfo{year}{1993}\natexlab{}.
\newblock \showarticletitle{Duration neglect in retrospective evaluations of
  affective episodes.}
\newblock \bibinfo{journal}{\emph{Journal of personality and social
  psychology}}  \bibinfo{volume}{65 1} (\bibinfo{year}{1993}),
  \bibinfo{pages}{45--55}.
\newblock


\bibitem[\protect\citeauthoryear{J\"{a}rvelin and
  Kek\"{a}l\"{a}inen}{J\"{a}rvelin and Kek\"{a}l\"{a}inen}{2002}]%
        {Jarvelin-2002}
\bibfield{author}{\bibinfo{person}{Kalervo J\"{a}rvelin} {and}
  \bibinfo{person}{Jaana Kek\"{a}l\"{a}inen}.} \bibinfo{year}{2002}\natexlab{}.
\newblock \showarticletitle{Cumulated Gain-Based Evaluation of IR Techniques}.
\newblock \bibinfo{journal}{\emph{ACM Trans. Inf. Syst.}} \bibinfo{volume}{20},
  \bibinfo{number}{4} (\bibinfo{date}{oct} \bibinfo{year}{2002}),
  \bibinfo{pages}{422–446}.
\newblock
\showISSN{1046-8188}
\urldef\tempurl%
\url{https://doi.org/10.1145/582415.582418}
\showDOI{\tempurl}


\bibitem[\protect\citeauthoryear{Kanoulas and Aslam}{Kanoulas and
  Aslam}{2009}]%
        {Kanoulas-2009}
\bibfield{author}{\bibinfo{person}{Evangelos Kanoulas} {and}
  \bibinfo{person}{Javed~A. Aslam}.} \bibinfo{year}{2009}\natexlab{}.
\newblock \showarticletitle{Empirical Justification of the Gain and Discount
  Function for NDCG}. In \bibinfo{booktitle}{\emph{Proceedings of the 18th ACM
  Conference on Information and Knowledge Management}} (Hong Kong, China)
  \emph{(\bibinfo{series}{CIKM '09})}. \bibinfo{publisher}{Association for
  Computing Machinery}, \bibinfo{address}{New York, NY, USA},
  \bibinfo{pages}{611–620}.
\newblock
\showISBNx{9781605585123}
\urldef\tempurl%
\url{https://doi.org/10.1145/1645953.1646032}
\showDOI{\tempurl}


\bibitem[\protect\citeauthoryear{Liu and Yu}{Liu and Yu}{2021}]%
        {Liu-2021-CIKM}
\bibfield{author}{\bibinfo{person}{Jiqun Liu} {and} \bibinfo{person}{Ran Yu}.}
  \bibinfo{year}{2021}\natexlab{}.
\newblock \bibinfo{booktitle}{\emph{State-Aware Meta-Evaluation of Evaluation
  Metrics in Interactive Information Retrieval}}.
\newblock \bibinfo{publisher}{Association for Computing Machinery},
  \bibinfo{address}{New York, NY, USA}, \bibinfo{pages}{3258–3262}.
\newblock
\showISBNx{9781450384469}
\urldef\tempurl%
\url{https://doi.org/10.1145/3459637.3482190}
\showURL{%
\tempurl}


\bibitem[\protect\citeauthoryear{Moffat, Bailey, Scholer, and Thomas}{Moffat
  et~al\mbox{.}}{2017}]%
        {Moffat-2017-Incorporating}
\bibfield{author}{\bibinfo{person}{Alistair Moffat}, \bibinfo{person}{Peter
  Bailey}, \bibinfo{person}{Falk Scholer}, {and} \bibinfo{person}{Paul
  Thomas}.} \bibinfo{year}{2017}\natexlab{}.
\newblock \showarticletitle{Incorporating User Expectations and Behavior into
  the Measurement of Search Effectiveness}.
\newblock \bibinfo{journal}{\emph{ACM Trans. Inf. Syst.}} \bibinfo{volume}{35},
  \bibinfo{number}{3}, Article \bibinfo{articleno}{24} (\bibinfo{date}{jun}
  \bibinfo{year}{2017}), \bibinfo{numpages}{38}~pages.
\newblock
\showISSN{1046-8188}
\urldef\tempurl%
\url{https://doi.org/10.1145/3052768}
\showDOI{\tempurl}


\bibitem[\protect\citeauthoryear{Moffat, Mackenzie, Thomas, and
  Azzopardi}{Moffat et~al\mbox{.}}{2022}]%
        {Moffat-2022}
\bibfield{author}{\bibinfo{person}{Alistair Moffat}, \bibinfo{person}{Joel
  Mackenzie}, \bibinfo{person}{Paul Thomas}, {and} \bibinfo{person}{Leif
  Azzopardi}.} \bibinfo{year}{2022}\natexlab{}.
\newblock \showarticletitle{A flexible framework for offline effectiveness
  metrics}. In \bibinfo{booktitle}{\emph{2022 International ACM SIGIR
  Conference on Research and Development in Information Retrieval}}.
\newblock


\bibitem[\protect\citeauthoryear{Moffat, Scholer, and Thomas}{Moffat
  et~al\mbox{.}}{2012}]%
        {Moffat-2012}
\bibfield{author}{\bibinfo{person}{Alistair Moffat}, \bibinfo{person}{Falk
  Scholer}, {and} \bibinfo{person}{Paul Thomas}.}
  \bibinfo{year}{2012}\natexlab{}.
\newblock \showarticletitle{Models and metrics: IR evaluation as a user
  process}. In \bibinfo{booktitle}{\emph{ADCS}}.
\newblock


\bibitem[\protect\citeauthoryear{Moffat, Thomas, and Scholer}{Moffat
  et~al\mbox{.}}{2013}]%
        {Moffat-2013-Users}
\bibfield{author}{\bibinfo{person}{Alistair Moffat}, \bibinfo{person}{Paul
  Thomas}, {and} \bibinfo{person}{Falk Scholer}.}
  \bibinfo{year}{2013}\natexlab{}.
\newblock \showarticletitle{Users versus Models: What Observation Tells Us
  about Effectiveness Metrics}. In \bibinfo{booktitle}{\emph{Proceedings of the
  22nd ACM International Conference on Information \& Knowledge Management}}
  (San Francisco, California, USA) \emph{(\bibinfo{series}{CIKM '13})}.
  \bibinfo{publisher}{Association for Computing Machinery},
  \bibinfo{address}{New York, NY, USA}, \bibinfo{pages}{659–668}.
\newblock
\showISBNx{9781450322638}
\urldef\tempurl%
\url{https://doi.org/10.1145/2505515.2507665}
\showDOI{\tempurl}


\bibitem[\protect\citeauthoryear{Moffat and Zobel}{Moffat and Zobel}{2008}]%
        {Moffat-2008}
\bibfield{author}{\bibinfo{person}{Alistair Moffat} {and}
  \bibinfo{person}{Justin Zobel}.} \bibinfo{year}{2008}\natexlab{}.
\newblock \showarticletitle{Rank-Biased Precision for Measurement of Retrieval
  Effectiveness}.
\newblock \bibinfo{journal}{\emph{ACM Trans. Inf. Syst.}} \bibinfo{volume}{27},
  \bibinfo{number}{1}, Article \bibinfo{articleno}{2} (\bibinfo{date}{dec}
  \bibinfo{year}{2008}), \bibinfo{numpages}{27}~pages.
\newblock
\showISSN{1046-8188}
\urldef\tempurl%
\url{https://doi.org/10.1145/1416950.1416952}
\showDOI{\tempurl}


\bibitem[\protect\citeauthoryear{Ravana and Moffat}{Ravana and Moffat}{2010}]%
        {ravana-2010}
\bibfield{author}{\bibinfo{person}{Sri~Devi Ravana} {and}
  \bibinfo{person}{Alistair Moffat}.} \bibinfo{year}{2010}\natexlab{}.
\newblock \showarticletitle{Score Estimation, Incomplete Judgments, and
  Significance Testing in IR Evaluation}. In
  \bibinfo{booktitle}{\emph{Information Retrieval Technology}},
  \bibfield{editor}{\bibinfo{person}{Pu-Jen Cheng}, \bibinfo{person}{Min-Yen
  Kan}, \bibinfo{person}{Wai Lam}, {and} \bibinfo{person}{Preslav Nakov}}
  (Eds.). \bibinfo{publisher}{Springer Berlin Heidelberg},
  \bibinfo{address}{Berlin, Heidelberg}, \bibinfo{pages}{97--109}.
\newblock


\bibitem[\protect\citeauthoryear{Robertson, Kanoulas, and Yilmaz}{Robertson
  et~al\mbox{.}}{2010}]%
        {robertson-2010}
\bibfield{author}{\bibinfo{person}{Stephen~E. Robertson},
  \bibinfo{person}{Evangelos Kanoulas}, {and} \bibinfo{person}{Emine Yilmaz}.}
  \bibinfo{year}{2010}\natexlab{}.
\newblock \showarticletitle{Extending Average Precision to Graded Relevance
  Judgments}. In \bibinfo{booktitle}{\emph{Proceedings of the 33rd
  International ACM SIGIR Conference on Research and Development in Information
  Retrieval}} (Geneva, Switzerland) \emph{(\bibinfo{series}{SIGIR '10})}.
  \bibinfo{publisher}{Association for Computing Machinery},
  \bibinfo{address}{New York, NY, USA}, \bibinfo{pages}{603–610}.
\newblock
\showISBNx{9781450301534}
\urldef\tempurl%
\url{https://doi.org/10.1145/1835449.1835550}
\showDOI{\tempurl}


\bibitem[\protect\citeauthoryear{Sakai}{Sakai}{2006}]%
        {Sakai-2006}
\bibfield{author}{\bibinfo{person}{Tetsuya Sakai}.}
  \bibinfo{year}{2006}\natexlab{}.
\newblock \showarticletitle{Evaluating Evaluation Metrics Based on the
  Bootstrap}. In \bibinfo{booktitle}{\emph{Proceedings of the 29th Annual
  International ACM SIGIR Conference on Research and Development in Information
  Retrieval}} (Seattle, Washington, USA) \emph{(\bibinfo{series}{SIGIR '06})}.
  \bibinfo{publisher}{Association for Computing Machinery},
  \bibinfo{address}{New York, NY, USA}, \bibinfo{pages}{525–532}.
\newblock
\showISBNx{1595933697}
\urldef\tempurl%
\url{https://doi.org/10.1145/1148170.1148261}
\showDOI{\tempurl}


\bibitem[\protect\citeauthoryear{Sakai}{Sakai}{2014}]%
        {Sakai-2014-promise}
\bibfield{author}{\bibinfo{person}{Tetsuya Sakai}.}
  \bibinfo{year}{2014}\natexlab{}.
\newblock \bibinfo{booktitle}{\emph{Metrics, Statistics, Tests}}.
\newblock \bibinfo{publisher}{Springer Berlin Heidelberg},
  \bibinfo{address}{Berlin, Heidelberg}, \bibinfo{pages}{116--163}.
\newblock
\urldef\tempurl%
\url{https://doi.org/10.1007/978-3-642-54798-0_6}
\showDOI{\tempurl}


\bibitem[\protect\citeauthoryear{Sakai}{Sakai}{2018}]%
        {Sakai-2018}
\bibfield{author}{\bibinfo{person}{Tetsuya Sakai}.}
  \bibinfo{year}{2018}\natexlab{}.
\newblock \showarticletitle{Laboratory experiments in information retrieval}.
\newblock \bibinfo{journal}{\emph{The information retrieval series}}
  \bibinfo{volume}{40} (\bibinfo{year}{2018}).
\newblock


\bibitem[\protect\citeauthoryear{Sakai}{Sakai}{2021}]%
        {Sakai-2021-ecir}
\bibfield{author}{\bibinfo{person}{Tetsuya Sakai}.}
  \bibinfo{year}{2021}\natexlab{}.
\newblock \showarticletitle{On the Instability of Diminishing Return IR
  Measures}. In \bibinfo{booktitle}{\emph{Advances in Information Retrieval}},
  \bibfield{editor}{\bibinfo{person}{Djoerd Hiemstra},
  \bibinfo{person}{Marie-Francine Moens}, \bibinfo{person}{Josiane Mothe},
  \bibinfo{person}{Raffaele Perego}, \bibinfo{person}{Martin Potthast}, {and}
  \bibinfo{person}{Fabrizio Sebastiani}} (Eds.). \bibinfo{publisher}{Springer
  International Publishing}, \bibinfo{address}{Cham},
  \bibinfo{pages}{572--586}.
\newblock
\showISBNx{978-3-030-72113-8}


\bibitem[\protect\citeauthoryear{Sakai, Tao, Zeng, Zheng, Mao, Chu, Liu,
  Maistro, Dou, Ferro, and Soboroff}{Sakai et~al\mbox{.}}{2021}]%
        {Sakai-2021-WWW3}
\bibfield{author}{\bibinfo{person}{Tetsuya Sakai}, \bibinfo{person}{Sijie Tao},
  \bibinfo{person}{Zhaohao Zeng}, \bibinfo{person}{Yukun Zheng},
  \bibinfo{person}{Jiaxin Mao}, \bibinfo{person}{Zhumin Chu},
  \bibinfo{person}{Yiqun Liu}, \bibinfo{person}{Maria Maistro},
  \bibinfo{person}{Zhicheng Dou}, \bibinfo{person}{Nicola Ferro}, {and}
  \bibinfo{person}{Ian Soboroff}.} \bibinfo{year}{2021}\natexlab{}.
\newblock \showarticletitle{Overview of the NTCIR-15 We Want Web with CENTRE
  (WWW-3) Task}.
\newblock


\bibitem[\protect\citeauthoryear{Sanderson}{Sanderson}{2010}]%
        {Sanderson-2010}
\bibfield{author}{\bibinfo{person}{Mark Sanderson}.}
  \bibinfo{year}{2010}\natexlab{}.
\newblock \showarticletitle{Test Collection Based Evaluation of Information
  Retrieval Systems}.
\newblock \bibinfo{journal}{\emph{Foundations and Trends in Information
  Retrieval}}  \bibinfo{volume}{4} (\bibinfo{date}{01} \bibinfo{year}{2010}),
  \bibinfo{pages}{247--375}.
\newblock
\urldef\tempurl%
\url{https://doi.org/10.1561/1500000009}
\showDOI{\tempurl}


\bibitem[\protect\citeauthoryear{Sanderson, Paramita, Clough, and
  Kanoulas}{Sanderson et~al\mbox{.}}{2010}]%
        {sanderson-2010-preference}
\bibfield{author}{\bibinfo{person}{Mark Sanderson},
  \bibinfo{person}{Monica~Lestari Paramita}, \bibinfo{person}{Paul Clough},
  {and} \bibinfo{person}{Evangelos Kanoulas}.} \bibinfo{year}{2010}\natexlab{}.
\newblock \showarticletitle{Do User Preferences and Evaluation Measures Line
  Up?}. In \bibinfo{booktitle}{\emph{Proceedings of the 33rd International ACM
  SIGIR Conference on Research and Development in Information Retrieval}}
  (Geneva, Switzerland) \emph{(\bibinfo{series}{SIGIR '10})}.
  \bibinfo{publisher}{Association for Computing Machinery},
  \bibinfo{address}{New York, NY, USA}, \bibinfo{pages}{555–562}.
\newblock
\showISBNx{9781450301534}
\urldef\tempurl%
\url{https://doi.org/10.1145/1835449.1835542}
\showDOI{\tempurl}


\bibitem[\protect\citeauthoryear{Sanderson and Zobel}{Sanderson and
  Zobel}{2005}]%
        {Sanderson-2005}
\bibfield{author}{\bibinfo{person}{Mark Sanderson} {and}
  \bibinfo{person}{Justin Zobel}.} \bibinfo{year}{2005}\natexlab{}.
\newblock \showarticletitle{Information Retrieval System Evaluation: Effort,
  Sensitivity, and Reliability}. In \bibinfo{booktitle}{\emph{Proceedings of
  the 28th Annual International ACM SIGIR Conference on Research and
  Development in Information Retrieval}} (Salvador, Brazil)
  \emph{(\bibinfo{series}{SIGIR '05})}. \bibinfo{publisher}{Association for
  Computing Machinery}, \bibinfo{address}{New York, NY, USA},
  \bibinfo{pages}{162–169}.
\newblock
\showISBNx{1595930345}
\urldef\tempurl%
\url{https://doi.org/10.1145/1076034.1076064}
\showDOI{\tempurl}


\bibitem[\protect\citeauthoryear{Voorhees}{Voorhees}{1998}]%
        {Voorhees-1998}
\bibfield{author}{\bibinfo{person}{Ellen~M. Voorhees}.}
  \bibinfo{year}{1998}\natexlab{}.
\newblock \showarticletitle{Variations in Relevance Judgments and the
  Measurement of Retrieval Effectiveness}. In
  \bibinfo{booktitle}{\emph{Proceedings of the 21st Annual International ACM
  SIGIR Conference on Research and Development in Information Retrieval}}
  (Melbourne, Australia) \emph{(\bibinfo{series}{SIGIR '98})}.
  \bibinfo{publisher}{Association for Computing Machinery},
  \bibinfo{address}{New York, NY, USA}, \bibinfo{pages}{315–323}.
\newblock
\showISBNx{1581130155}
\urldef\tempurl%
\url{https://doi.org/10.1145/290941.291017}
\showDOI{\tempurl}


\bibitem[\protect\citeauthoryear{Voorhees}{Voorhees}{2002}]%
        {Voorhees-2002}
\bibfield{author}{\bibinfo{person}{Ellen~M. Voorhees}.}
  \bibinfo{year}{2002}\natexlab{}.
\newblock \showarticletitle{The Philosophy of Information Retrieval
  Evaluation}. In \bibinfo{booktitle}{\emph{Evaluation of Cross-Language
  Information Retrieval Systems}}, \bibfield{editor}{\bibinfo{person}{Carol
  Peters}, \bibinfo{person}{Martin Braschler}, \bibinfo{person}{Julio Gonzalo},
  {and} \bibinfo{person}{Michael Kluck}} (Eds.). \bibinfo{publisher}{Springer
  Berlin Heidelberg}, \bibinfo{address}{Berlin, Heidelberg},
  \bibinfo{pages}{355--370}.
\newblock


\bibitem[\protect\citeauthoryear{Voorhees}{Voorhees}{2009}]%
        {Voorhees-2009}
\bibfield{author}{\bibinfo{person}{Ellen~M. Voorhees}.}
  \bibinfo{year}{2009}\natexlab{}.
\newblock \showarticletitle{Topic Set Size Redux}. In
  \bibinfo{booktitle}{\emph{Proceedings of the 32nd International ACM SIGIR
  Conference on Research and Development in Information Retrieval}} (Boston,
  MA, USA) \emph{(\bibinfo{series}{SIGIR '09})}.
  \bibinfo{publisher}{Association for Computing Machinery},
  \bibinfo{address}{New York, NY, USA}, \bibinfo{pages}{806–807}.
\newblock
\showISBNx{9781605584836}
\urldef\tempurl%
\url{https://doi.org/10.1145/1571941.1572138}
\showDOI{\tempurl}


\bibitem[\protect\citeauthoryear{Voorhees and Buckley}{Voorhees and
  Buckley}{2002}]%
        {Voorhees-2002-effect}
\bibfield{author}{\bibinfo{person}{Ellen~M. Voorhees} {and}
  \bibinfo{person}{Chris Buckley}.} \bibinfo{year}{2002}\natexlab{}.
\newblock \showarticletitle{The Effect of Topic Set Size on Retrieval
  Experiment Error}. In \bibinfo{booktitle}{\emph{Proceedings of the 25th
  Annual International ACM SIGIR Conference on Research and Development in
  Information Retrieval}} (Tampere, Finland) \emph{(\bibinfo{series}{SIGIR
  '02})}. \bibinfo{publisher}{Association for Computing Machinery},
  \bibinfo{address}{New York, NY, USA}, \bibinfo{pages}{316–323}.
\newblock
\showISBNx{1581135610}
\urldef\tempurl%
\url{https://doi.org/10.1145/564376.564432}
\showDOI{\tempurl}


\bibitem[\protect\citeauthoryear{Wicaksono and Moffat}{Wicaksono and
  Moffat}{2020}]%
        {Wicaksono-2020-Metrics}
\bibfield{author}{\bibinfo{person}{Alfan~Farizki Wicaksono} {and}
  \bibinfo{person}{Alistair Moffat}.} \bibinfo{year}{2020}\natexlab{}.
\newblock \showarticletitle{Metrics, User Models, and Satisfaction}. In
  \bibinfo{booktitle}{\emph{Proceedings of the 13th International Conference on
  Web Search and Data Mining}} (Houston, TX, USA) \emph{(\bibinfo{series}{WSDM
  '20})}. \bibinfo{publisher}{Association for Computing Machinery},
  \bibinfo{address}{New York, NY, USA}, \bibinfo{pages}{654–662}.
\newblock
\showISBNx{9781450368223}
\urldef\tempurl%
\url{https://doi.org/10.1145/3336191.3371799}
\showDOI{\tempurl}


\bibitem[\protect\citeauthoryear{Zhang, Liu, Li, Zhang, Xu, and Ma}{Zhang
  et~al\mbox{.}}{2017}]%
        {Zhang-2017}
\bibfield{author}{\bibinfo{person}{Fan Zhang}, \bibinfo{person}{Yiqun Liu},
  \bibinfo{person}{Xin Li}, \bibinfo{person}{Min Zhang},
  \bibinfo{person}{Yinghui Xu}, {and} \bibinfo{person}{Shaoping Ma}.}
  \bibinfo{year}{2017}\natexlab{}.
\newblock \showarticletitle{Evaluating Web Search with a Bejeweled Player
  Model}. In \bibinfo{booktitle}{\emph{Proceedings of the 40th International
  ACM SIGIR Conference on Research and Development in Information Retrieval}}
  (Shinjuku, Tokyo, Japan) \emph{(\bibinfo{series}{SIGIR '17})}.
  \bibinfo{publisher}{Association for Computing Machinery},
  \bibinfo{address}{New York, NY, USA}, \bibinfo{pages}{425–434}.
\newblock
\showISBNx{9781450350228}
\urldef\tempurl%
\url{https://doi.org/10.1145/3077136.3080841}
\showDOI{\tempurl}


\bibitem[\protect\citeauthoryear{Zhang, Mao, Liu, Xie, Ma, Zhang, and Ma}{Zhang
  et~al\mbox{.}}{2020}]%
        {Zhang-2020-Models}
\bibfield{author}{\bibinfo{person}{Fan Zhang}, \bibinfo{person}{Jiaxin Mao},
  \bibinfo{person}{Yiqun Liu}, \bibinfo{person}{Xiaohui Xie},
  \bibinfo{person}{Weizhi Ma}, \bibinfo{person}{Min Zhang}, {and}
  \bibinfo{person}{Shaoping Ma}.} \bibinfo{year}{2020}\natexlab{}.
\newblock \showarticletitle{Models Versus Satisfaction: Towards a Better
  Understanding of Evaluation Metrics}. In
  \bibinfo{booktitle}{\emph{Proceedings of the 43rd International ACM SIGIR
  Conference on Research and Development in Information Retrieval}} (Virtual
  Event, China) \emph{(\bibinfo{series}{SIGIR '20})}.
  \bibinfo{publisher}{Association for Computing Machinery},
  \bibinfo{address}{New York, NY, USA}, \bibinfo{pages}{379–388}.
\newblock
\showISBNx{9781450380164}
\urldef\tempurl%
\url{https://doi.org/10.1145/3397271.3401162}
\showDOI{\tempurl}


\bibitem[\protect\citeauthoryear{Zobel}{Zobel}{1998}]%
        {Zobel-1998}
\bibfield{author}{\bibinfo{person}{Justin Zobel}.}
  \bibinfo{year}{1998}\natexlab{}.
\newblock \showarticletitle{How Reliable Are the Results of Large-Scale
  Information Retrieval Experiments?}. In \bibinfo{booktitle}{\emph{Proceedings
  of the 21st Annual International ACM SIGIR Conference on Research and
  Development in Information Retrieval}} (Melbourne, Australia)
  \emph{(\bibinfo{series}{SIGIR '98})}. \bibinfo{publisher}{Association for
  Computing Machinery}, \bibinfo{address}{New York, NY, USA},
  \bibinfo{pages}{307–314}.
\newblock
\showISBNx{1581130155}
\urldef\tempurl%
\url{https://doi.org/10.1145/290941.291014}
\showDOI{\tempurl}


\end{thebibliography}
\end{document}